\documentclass[prl,aps,reprint,footnotebib,amssymb]{revtex4-2}
\usepackage{amsmath}
\usepackage{bm}
\usepackage[dvipsnames]{xcolor}
\usepackage{graphicx}
\usepackage{epstopdf}
\usepackage{epsfig}
\usepackage{amsfonts}
\usepackage[pdfencoding=auto,naturalnames]{hyperref}
\usepackage{hypcap}
\usepackage{mathtools}
\usepackage{verbatim}
\usepackage{tabularx}
\usepackage{bbm}
\usepackage{esvect}
\usepackage{subfigure}
\usepackage{slashed}
\usepackage{physics}
\usepackage{times}
\usepackage[normalem]{ulem}
\usepackage{orcidlink}
\usepackage{array}
\usepackage{multirow}

\def\bea{\begin{eqnarray}}
\def\eea{\end{eqnarray}}
\def\be{\begin{equation}}
\def\ee{\end{equation}}

\def\ba{\begin{array}}
\def\ea{\end{array}}
\def\Tr{\mathrm{Tr}}
\DeclareMathOperator{\sgn}{sgn}

\newcommand{\mathd}{\mathrm{d}}

\newcommand{\tmmathbf}[1]{\ensuremath{\boldsymbol{#1}}}
\newcommand{\tmop}[1]{\ensuremath{\operatorname{#1}}}

\newcommand{\ssll}{\hspace{-1em}---}

\newcommand{\rrangle}{\ensuremath{\rangle\hskip-0.2em\rangle}}

\hypersetup{colorlinks=true, citecolor=Blue, urlcolor=Blue, linkcolor=Blue, breaklinks=true, pdfduplex=DuplexFlipLongEdge, %
pdftitle={Defect Conformal Field Theory from Sachdev-Ye-Kitaev Interactions}, %
pdfauthor={Yang Ge, Shaokai Jian}}

\begin{document}
\title{Defect Conformal Field Theory from Sachdev-Ye-Kitaev Interactions}

\author{Yang Ge\,\orcidlink{0000-0003-4866-5899}}
\author{Shao-Kai Jian\,\orcidlink{0000-0003-3058-8725}}
\email{sjian@tulane.edu}
\affiliation{Department of Physics and Engineering Physics, Tulane University, New Orleans, Louisiana 70118, USA}

\date{\today}

\begin{abstract}
    The coupling between defects and extended critical degrees of freedom gives rise to the intriguing theory known as defect conformal field theory (CFT). 
    In this work, we introduce a novel family of boundary and interface CFTs by coupling $N$ Majorana chains with SYK$_q$ interactions at the defect. 
    Our analysis reveals that the interaction with $q=2$ constitutes a new marginal defect. 
    Employing a versatile saddle-point method, we compute unique entanglement characterizations, including the $g$ function and effective central charge, of the defect CFT. 
    Furthermore, we analytically evaluate the transmission coefficient using CFT techniques. 
    Surprisingly, the transmission coefficient deviates from the universal relation with the effective central charge across the defect at the large $N$ limit, suggesting that our defect CFT extends beyond all known examples of Gaussian defect CFT.
\end{abstract}

\maketitle

\paragraph{Introduction}\ssll
Understanding the defect or boundary conformal field theory (CFT) holds significant implications across various domains of theoretical physics~\cite{cardy1984conformal,cardy1989boundary,andrei2020boundary}. 
In condensed matter physics, defect CFT (dCFT) provides a powerful framework for deciphering critical behaviors of complex materials characterized by boundaries, interfaces, and defects, all common in the real world.  
In particular, boundary phenomena host the most interesting physics in symmetry-protected topological phases~\cite{hasan2010topological,qi2011topological}.
Within the framework of string theory, dCFT naturally emerges in the study of D-branes~\cite{recknagel2013boundary,polchinski1996tasi}, offering insights on topics including brane intersections and holographic correspondences~\cite{maldacena1999large,witten1998anti,gubser1998gauge} between gravitational theories and boundary CFTs~\cite{takayanagi2011holographic,fujita2011aspects}.

Transmission and reflection are important characterizations of interfaces in CFTs~\cite{kane1992transmission,quella2007reflection}. 
Interactions can render defects relevant or irrelevant, leading to asymptotic behaviors where defects become completely reflective or transmitting. 
In 2D free massless fermion theories, it was discovered that defects can be marginal, resulting in partial transmission and reflection~\cite{oshikawa1996defect,oshikawa1997boundary,delfino1994scattering,quella2007reflection,bachas2002permeable,bachas2013fusion}.
Beyond free dCFTs, non-Gaussian defects are only understood on a case-by-case basis. Two special cases are the totally reflective factorizing defect and the perfectly transmitting topological defect in rational CFTs \cite{frohlich2004kramers,frohlich2007duality}. Defects other than total transmitters or reflectors in two dimensions include the symmetry breaking defect via a coset construction \cite{quella2002symmetry}, permutationlike boundary conditions in Wess-Zumino-Witten and coset models \cite{fredenhagen2005generalised,brunner2005matrix}, defects in the tricritical Ising model \cite{makabe2017tcim}, as well as in minimal models with rational products \cite{quella2007reflection} or random defects~\cite{jeng2001random}.
Higher-dimensional cases include defects in the 3D Ising model \cite{hu2024solving}, O($N$) model \cite{metlitski2022boundary,parisen-toldin2022boundary,krishnan2023plane,giombi2023notes,trepanier2023surface,raviv-moshe2023phases},  Super-Yang-Mills theories \cite{constable2000noncommutative,karch2001locally,dewolfe2002holography,erdmenger2002scft}, the AdS/dCFT construction \cite{takayanagi2011holographic,fujita2011aspects}, as well as the Kondo model~\cite{affleck1991kondo}. 

Entanglement emerges as a useful characterization of many-body wave functions~\cite{amico2008entanglement,horodecki2009quantum,calabrese2009entanglement1,eisert2010area}. 
Many-body entanglement can be quantified with the celebrated R\'enyi entropy \cite{jizba2004renyi}: given a many-body wave function $|\Psi \rangle$ and a bipartition $A \cup B$, the R\'enyi entropy is $S_n(A)= \frac1{1-n} \log \Tr[\rho_A^n]$, with $\rho_A = \Tr_{B}[\rho]$ being the reduced density matrix on $A$.
While the entanglement entropy in a CFT is closely related to its central charge~\cite{calabrese2004entanglement,calabrese2009entanglement,blote1986conformal,affleck1986universal}, it can be altered by defects.
In this context, the boundary entropy, or $g$ function~\cite{affleck1991universal}, universally characterizes the ground state degeneracy in the presence of boundaries or defects.
More explicitly, boundary conditions contribute to the free energy by a constant independent of its system size $L$ when $L$ is large. 
Additionally, for marginal defects in free fermion CFTs, the R\'enyi entropy across the defect is captured by an effective central charge, which exhibits a universal function of the transmission coefficient~\cite{sakai2008entanglement,eisler2010solution,calabrese2012entanglement,peschel2012exact,brehm2015entanglement,capizzi2022renyi,capizzi2022renyiboson,rogerson2022entanglement,roy2022entanglement}.

The Sachdev-Ye-Kitaev (SYK) model \cite{sachdev1993gapless,kitaev2015simple,maldacena2016remarks,polchinski2016spectrum}, a 0+1D quantum model, presents a unique and solvable candidate for defects beyond all known examples. 
Initially introduced as a solvable toy model, with intriguing properties akin to black holes, the SYK model has found solvable generalizations in various fields, including non-Fermi liquid behavior~\cite{chowdhury2022sachdev,song2017strongly,bi2017instability,jian2018quantum,chowdhury2018translationally,patel2018magnetotransport}, thermalization~\cite{you2017thermalization,jian2017solvable,sonner2017eigenstate,gu2017spread,garcia2018chaotic,liu2018quantum,dai2019global}, and non-Hermitian physics~\cite{liu2021non,zhang2021emergent, jian2021measurement,garcia2022replica,garcia2022dominance}. 
The interplay between the SYK model and CFTs has been studied in the context of black hole evaporation~\cite{penington2022replica,almheiri2020replica,chen2020replica,liu2022black}. 
However, the joint system, where SYK acts as a defect~\cite{shen2023long,gao2023information,gao2024scrambling}, remains relatively unexplored. 
Key questions include constructing a dCFT from the SYK model and identifying unique characterizations of such a dCFT.

\begin{figure}
    \includegraphics[width=0.7\linewidth]{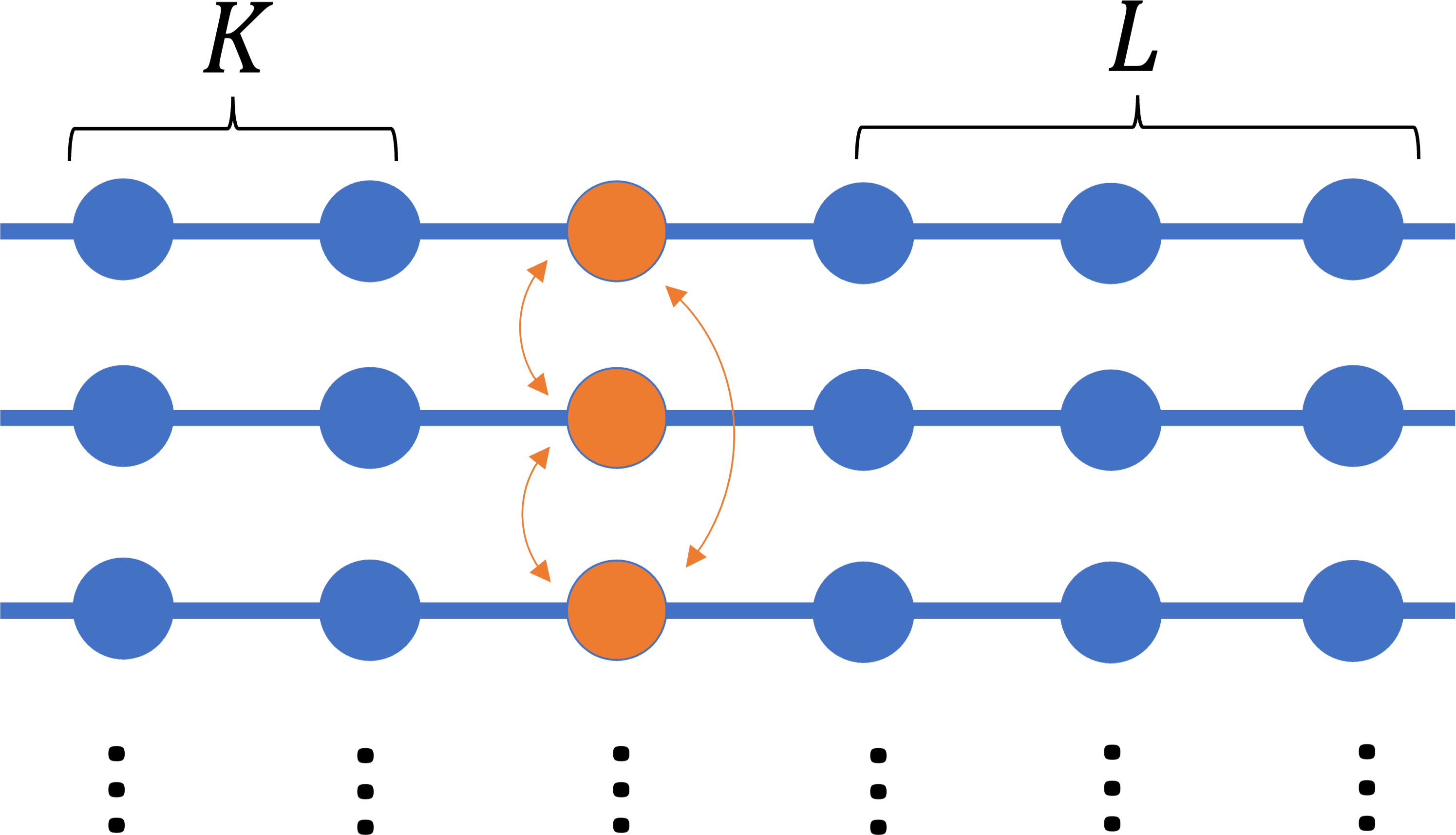}
    \caption{Illustration of the lattice model.
        The dots denote each site of the Majorana chain.
        There are $N$ free Majorana chains, that are coupled via SYK interactions at the orange site, dubbed the SYK site.
        Open boundary condition is assumed and the number of sites to the left (right) of the SYK site is $K$ ($L$).}
    \label{fig:model}
\end{figure}

In this Letter, we build boundary and defect CFTs by coupling $N$ Majorana chains with SYK$_q$-type interactions at the defect, illustrated in Fig.~\ref{fig:model}. 
We show that the interaction is marginal for $q=2$, and irrelevant for $q>2$.
For $q=2$, the analytical conformal solution gives a tunable transmission coefficient.
We develop a saddle-point method to investigate the R\'enyi entropy of various bipartitions in the joint system.
In an islandlike bipartition, the $g$ function is shown to be 1. 
Finally, we compute the R\'enyi entropy across the SYK defect and extract a continuous effective central charge on the SYK interaction strength. 
Unlike known Gaussian dCFTs, the universal relation between the effective central charge and the transmission breaks down for the SYK$_2$ defect \cite{capizzi2022renyi}, owing to the non-Gaussianity induced by the random coupling in SYK.

\paragraph{Model}\ssll
We consider the Hamiltonian $H=H_\text{CFT} + H_\text{I}$, where $H_\text{CFT} =  -i 2 t \sum_{jr} \psi_{j,r} \psi_{j,r+1}$ consists of $j\!=\!1 , \dots , N$ decoupled Majorana chains with nearest-neighbor hoppings of amplitude $t$ between sites $r$ and $r+1$, while $H_\text{I} = i^{q/2} \sum_{j_1,\dots, j_q}  J_{j_1,\dots,j_q} \psi_{j_1,0} \dots \psi_{j_q,0}$ couples the chains by the SYK$_q$ interaction of strength $J$ at the site $r \! = \! 0$, dubbed the SYK site.
Thus, each site has $N$ Majorana fermions so that
$\{\psi_{j,r}, \psi_{j',r'}\} = \delta_{j,j'} \delta_{r,r'}$.
The interaction strength $J_{j_1,\dots,j_q}$ is a Gaussian variable with mean zero and variance
$\overline{J_{j_1,\dots,j_q} J_{j'_1,\dots,j'_q}} = \delta_{j_1, j'_1} \dots \delta_{j_q, j'_q} \frac{2^{q-1} J^2 (q-1)!}{N^{q-1}}$.
Figure~\ref{fig:model} shows the schematic. 

\begin{figure}
    \includegraphics[width=\linewidth]{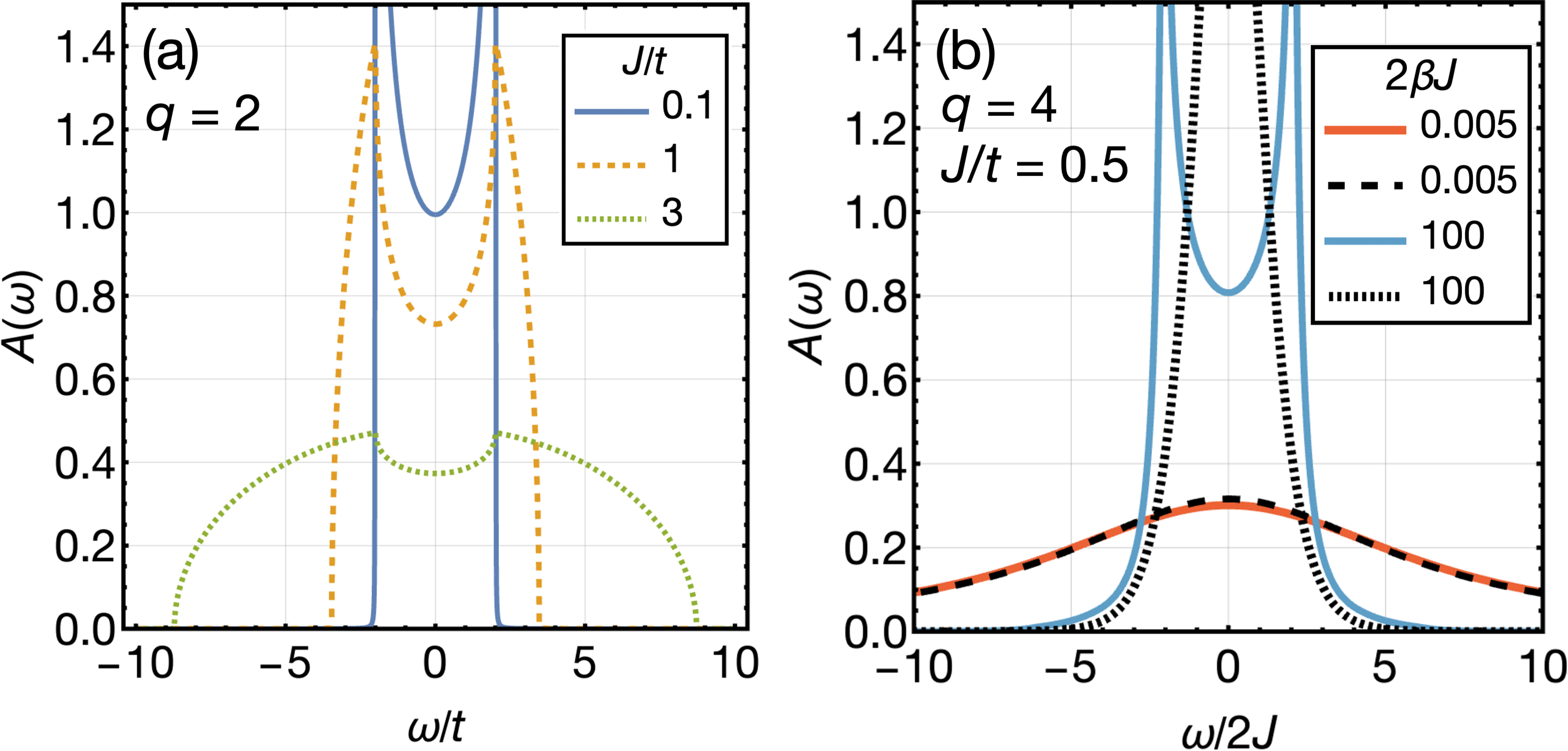}
    \caption{Local spectral functions at the SYK$_q$ site on infinite chains.
    (a)~The SYK$_2$-site spectral function at various $J/t$, independent of temperature.
    (b)~Temperature evolution of the spectral function at the SYK$_4$ site with $J/t=0.5$. Dashed curves show the spectral functions of the bare SYK$_4$ model. 
    Deviation from the SYK$_4$ fixed point at low temperature near zero frequency shows the irrelevance. 
    The spectral function flow is settled at $2\beta J\sim0.2$.}
    \label{fig:infinite}
\end{figure}

We use the $(G,\Sigma)$-action at the SYK site to solve the model \cite{maldacena2016remarks,bagrets2016SYK,gu2017spread,kitaev2018soft}, with 
$G(\tau) = \frac1N \sum_j \langle \psi_{j,0}(\tau) \psi_{j,0}(0) \rangle$, and $\Sigma$ being the corresponding self-energy \footnote{See Supplemental Material for further details.}.
Here the angle bracket denotes both quantum and disorder averages, and $\tau$ is imaginary time. 
In the large-$N$ limit, the Schwinger-Dyson equations are
\begin{eqnarray}
    G^{-1}(i\omega_n) &=& -i\omega_n - \Sigma(i\omega_n) - t R_{K}\!\left(\tfrac{-i\omega_n}{2t}\right)- t R_{L}\!\left(\tfrac{-i\omega_n}{2t}\right) \!,  \nonumber \\
    \Sigma(\tau) &=& J^2 [2G(\tau)]^{q-1}, \label{eq:dyson}
\end{eqnarray}
where $G(i \omega_n) \equiv \int d\tau G(\tau) e^{i\omega_n \tau}$ with the Matsubara frequencies $\omega_n \equiv (2n+1)\pi/\beta$ at the inverse temperature $\beta$. 
Coupling to the rest of the chains results in the self-energy terms of the form $R_L(x)\!\equiv\!\frac{U_{L-1}(x)}{U_L(x)}$, where $U$ is the Chebyshev polynomial of the second kind~\cite{mallik2001inverse,zhang2013matrix}.

A simple scaling analysis reveals that the SYK interaction is marginal for $q \! = \! 2$ and irrelevant for $q \! >\!2$, since the scaling dimension of $J$ is $1-\frac{q}{2}$.
This is confirmed by exactly solving the model in the thermodynamic limit, $K,L\!\to\!\infty$, using $\lim_{L \rightarrow \infty} R_L(x) = x - \sqrt{x^2 - 1}$.
The local spectral function at the SYK site, $A(\omega)\equiv 2G^{\prime\prime}(\omega-i\eta)$ is used to expand $G$ and express $\Sigma$~\cite{komijani2018ferro,Note1}.
Thus, $A(\omega)$ is obtained exactly and plotted in Fig.~\ref{fig:infinite}.
For $q \! = \! 2$, the interaction is indeed marginal and $A(\omega)$ temperature independent. Across all $J/t$ in Fig.~\ref{fig:infinite}(a), it features two peaks coming from the band edges of the Majorana chain, and a semicircle coming from the SYK$_2$ site. On the other hand, $A(\omega)$ is temperature dependent when $q \! > \! 2$. Figure~\ref{fig:infinite}(b) shows the example of the $q \! = \! 4$ case along with that of a decoupled SYK$_4$ site. At low temperature and near zero frequency, $A(\omega)$ flows away from the SYK fixed point, demonstrating its irrelevance.
Consequently, we will focus on $q \! = \! 2$ from now on.

In the SYK$_2$ case, the full Green's function on the chains takes a simple analytic form in the thermodynamic limit~\cite{Note1}, which can be derived from the Dyson equation that consists of the Green's function of the free Majorana chain $G_0$ and the self energy $\Sigma$ at the SYK site,
$	G (r, r' ; i\omega_n) =
	 G_0  (r, r' ;i \omega_n)+ 
	 G_0  (r, 0 ; i\omega_n) \Sigma (i\omega_n) G (0, r' ; i\omega_n). %
$
To further study the low-energy physics we focus on the %
the left- and right-propagating chiral modes at the Fermi surface, $k_F\!=\!0$ and $\pi$, with the lattice constant set to 1. They are $\psi_{L,R}(x)\sim\int_{k\sim 0,\pi} e^{-i k x} \psi(k) \mathd k$. Together they form a Dirac spinor $\Psi\! = \!(\psi_L,\psi_R)^T$.
After coarse graining~\cite{Note1},  the effective action for the SYK$_2$ model becomes a dCFT:
\begin{equation}
	\mathcal L_{\text{dCFT}} \!=\! \sum_j \Psi_j^\dag (\partial_\tau -  i t \partial_x \sigma^z) \Psi_j + \sum_{jl} i \delta(x) \tilde{J}_{jl} \Psi_j^{\dagger} P \Psi_l, \label{eq:LE}
\end{equation}
where $\tilde{J} \equiv J/t$ and $P = \frac12 \begin{psmallmatrix}
	1 & 1 \\
	1 & 1
\end{psmallmatrix}.$
The continuum Green's function can be solved explicitly to reveal its conformal nature~\cite{Note1}. By conformal symmetry, higher-order correlators recursively reduce to the single-particle Green's function~\cite{delfino1994scattering}.

\paragraph{Transmission and reflection}\ssll
A dCFT is characterized by its transmission and reflection coefficients, $\mathcal{T}$ and $\mathcal{R} \!=\! 1- \mathcal T$, across the defect \cite{delfino1994scattering,quella2007reflection}. They are the corresponding probabilities in single-particle free field theories. For a conformal defect at $x=0$, the transmission coefficient is defined by the holomorphic and antiholomorphic components of the stress-energy tensor, $T$ and $\bar{T}$, at two sides of the defect \cite{quella2007reflection},
\begin{equation}
	\mathcal T \equiv \frac{\langle  T(x) T(-x) + \bar T(x) \bar T(-x) \rangle }{ \langle [T(x) + \bar{T}(-x)] [\bar{T}(x) + T(-x)] \rangle} . \label{eq:Tdef}
\end{equation}
For the low energy CFT of the SYK$_2$ defect~\cite{Note1}
\begin{equation}
	\mathcal T %
	= \frac2{3+\tilde J^2 - \sqrt{1+2 \tilde J^2}}. %
	\label{eq:T}
\end{equation}
In general, coarse graining need not produce a simple relation between the lattice parameters at UV and continuum parameters at IR. Still, solutions to both the lattice and the continuum model \eqref{eq:LE} produce the same transmission coefficient~\cite{Note1}, with $\tilde{J}^2\to J^2 /(t^2 \cos^2 k)$. Thus, $\tilde{J}= J/t$ holds at long wavelengths for all $J/t$. At small $J$, $1-\mathcal{T} \sim \tilde{J}^4/4$. Our defect formally corresponds to the Dirichlet boundary condition in the Ising CFT, that is $|  D,\phi \rrangle$ with $\phi = \frac{1}{2}\arcsin \sqrt{\mathcal{T}}$~\cite{oshikawa1997boundary,brehm2015entanglement}.
As for SYK$_q$ defects with $q>2$, $\mathcal{T}=1$ due to their irrelevance.

\begin{figure}
    \includegraphics[width=\linewidth]{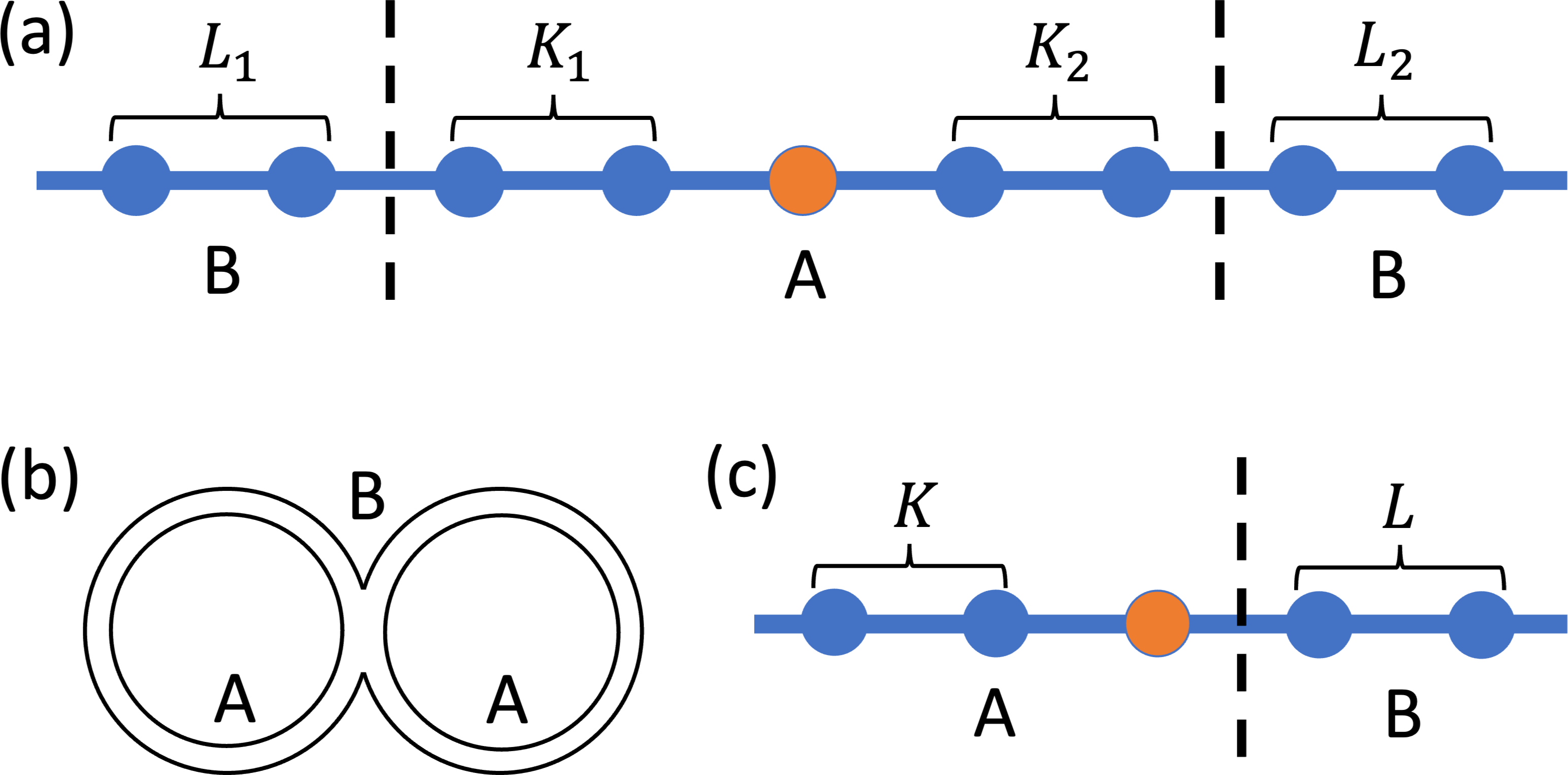}
    \caption{(a) Generic bipartition of 1+1D open chains. For simplicity, a single chain is plotted, which nevertheless should be understood as $N$ chains.
        (b) The imaginary time contour for the second R\'enyi entropy. The fermionic field in a closed contour satisfies the conventional antiperiodic boundary condition associated with the fermionic Matsubara frequency. Region $B$ has the twisted boundary condition threading the two replicas.
        (c) Setup for the interface CFT induced at the SYK site, where $L_1\!=\!K_2\!=\!0$.
    }
    \label{fig:renyi-contour}
\end{figure}

\paragraph{R\'enyi entropy}\ssll
Sitting between two 1+1D CFTs, the SYK defect encodes the entanglement across it in the conformal data of its dCFT \cite{sakai2008entanglement,brehm2015entanglement,capizzi2022renyi}.
Using the replica trick and the large-$N$ analysis, we present a method~\cite{liu2018quantum,chen2020replica,shao2024towards} to conveniently extract the R\'enyi entanglement entropy for an arbitrary bipartition as depicted in Fig.~\ref{fig:renyi-contour}(a). %
Below, the second R'enyi entropy is computed as an example and later used to obtain the $g$ function as well as the effective central charge.

The second R\'enyi entropy of region $A$ is given by partial traces over the reduced density matrix,
\begin{eqnarray}
    S_2 = - \log \Tr_A \! \left[ (\Tr_B \rho )^2 \right] = -\log \left[Z_{(2)}/Z^2 \right].
    \label{eq:S2}
\end{eqnarray}
The replica trick conducts the partial trace with different imaginary-time boundary conditions in regions $A$ and $B$ \cite{calabrese2004entanglement,liu2018quantum,chen2020replica}. 
Here $Z_{(2)}$ denotes a two-replica path integral with a twisted boundary condition in region $B$, while $Z$ is the path integral corresponding to the thermal density matrix $\rho \!\equiv\! e^{-\beta H}/Z$.
Boundary conditions of the replica fields differ between regions $A$ and $B$: in region $A$,
$\psi^{(1)}( \beta ) = -\psi^{(1)}( 0 )$, $\psi^{(2)}( \beta ) = - \psi^{(2)}( 0 )$, whereas in region $B$,
$\psi^{(1)}( \beta ) = \psi^{(2)}( 0 )$, $\psi^{(2)}( \beta ) = - \psi^{(1)}( 0 )$. The superscripts 1 and 2 denote the two replicas.
This imaginary time contour for $Z_{(2)}$ is depicted in Fig.~\ref{fig:renyi-contour}(b). 
We join the two replicas to span $\tau\in (0,2\beta)$, so that fields can be expanded in Matsubara frequencies $\omega^{(1)}_n \!=\! \omega^{(2)}_n =(2n+1)\pi/\beta$ in region $A$, and $\Omega_n = (2n+1)\pi/2\beta$ in region $B$. Namely, in region $A$,
$\psi (\tau) = \frac{1}{\sqrt{\beta}} \sum_n \psi (i \omega_n^{(1)}) e^{- i \omega^{(1)}_n \tau} \Theta(\beta-\tau) + \frac{1}{\sqrt{\beta}} \sum_n  \psi (i \omega_n^{(2)}) e^{- i \omega^{(2)}_n \tau} \Theta(\tau-\beta)$, whereas in region $B$,
$\psi (\tau) = \frac{1}{\sqrt{2 \beta}} \sum_n \psi (i \Omega_n) e^{- i \Omega_n \tau}$.

At large $N$, we solve for the saddle-point Green's function $G_{(2)}$ at the SYK site, and then substitute it into the twisted action to compute $Z_{(2)}$ \cite{Note1}.
The time-translation symmetry is broken in $Z_{(2)}$ due to the twist operator $\sigma$ at the partition interface, which is defined below.
The solution is
\begin{eqnarray}
    G^{-1}_{(2)}(i\omega^{(a)}_m,i\omega^{(b)}_n) &=& -i\omega^{(a)}_n \delta_{ab}\delta_{mn} - \Sigma_{(2)}(i\omega^{(a)}_m,i\omega^{(b)}_n)   \nonumber \\
    && - t (\mathbf{D}_{K_1,L_1})^{ab}_{mn} - t (\mathbf{D}_{K_2,L_2})^{ab}_{mn},   \nonumber \\
    \Sigma_{(2)}(\tau_1, \tau_2) &=& J^2 [2G_{(2)}(\tau_1, \tau_2)]^{q-1}. \label{eq:dyson2}
\end{eqnarray}
Here the self-energy due to the rest of the chain is $\mathbf{D}_{K,L}$~\cite{Note1}, defined recursively by
$\mathbf{D}_{0,L} = \sigma R_{L} \! \left( \tfrac{- i\boldsymbol{\Omega}}{2 t} \right) \sigma^{\dagger}$, and
$\mathbf{D}_{K,L} = \left( - \tfrac{i\boldsymbol{\omega}}{t} - \mathbf{D}_{K - 1,L} \right)^{-1}$,
using the Matsubara frequency matrices
$\boldsymbol{\omega}=\mathrm{diag}(\omega^{(1)}_1,\omega^{(1)}_2,\dots,\omega^{(2)}_1,\omega^{(2)}_2,\dots)$, and
$\boldsymbol{\Omega}=\mathrm{diag}(\Omega_1,\Omega_{2},\dots)$.
The twist operator $\sigma$ transforms fields across the partition. 
In the frequency space,
$\sigma (i \omega_m^{(a)}, i \Omega_n) = \int^{a\beta}_{(a-1)\beta} \frac{\mathd \tau}{\sqrt{2} \beta} e^{i [\omega_m^{(a)} - \Omega_n] \tau}$, for replicas $a=1,2$.

Following \eqref{eq:S2},  $S_2=2 \log Z -\log Z_{(2)}$ gives the second R\'enyi entropy.
The ground-state entanglement simplifies after subtracting the zero $S_2$ when $A$ and $B$ are decoupled. 
Denote the partition function in a decoupled system by $\check{Z}$, we get %
\begin{eqnarray}
\label{eq:S2island}
    \frac{S_2}{N} &=& \frac{1}{N} \left[ 2 \log Z -\log Z_{(2)} -\left( 2 \log \check{Z} -\log \check{Z}_{(2)} \right) \right] \\
    & = & - \frac{1}{2} \Tr \log [G_{(2)}^{- 1} \tilde{G}] + J^2  \left( \frac{1}{4 q} - \frac{1}{4} \right)  \nonumber\\
    &&\times \int \mathd \tau_1 \mathd \tau_2  \{ [2 \tilde{G} (\tau_1, \tau_2)]^q - [2 G_{(2)} (\tau_1, \tau_2)]^q \} \nonumber\\
    &  & + \sum_{s=1,2} \frac{1}{2} \Tr \log \left[ \frac{1 - R_{L_s} \! \left( \frac{- i\boldsymbol{\omega}}{2 t} \right) R_{K_s} \! \left( \frac{- i\boldsymbol{\omega}}{2 t} \right)}{1 - \sigma R_{L_s} \! \left( \frac{- i\boldsymbol{\Omega}}{2 t} \right) \sigma^{\dagger} R_{K_s} \! \left( \frac{- i\boldsymbol{\omega}}{2 t} \right)} \right] \!,  \nonumber
\end{eqnarray}
where $\tilde{G}=G\otimes\mathbf{1}_2$ is the Green's function for $Z$, i.e.\ \eqref{eq:dyson} with $K\to L_1+K_1$ and $L\to L_2+K_2$, replicated diagonally in the replica space. The last line is the entanglement between free Majorana chains with uniform nearest-neighbor hoppings.

\begin{figure}
    \includegraphics[width=\linewidth]{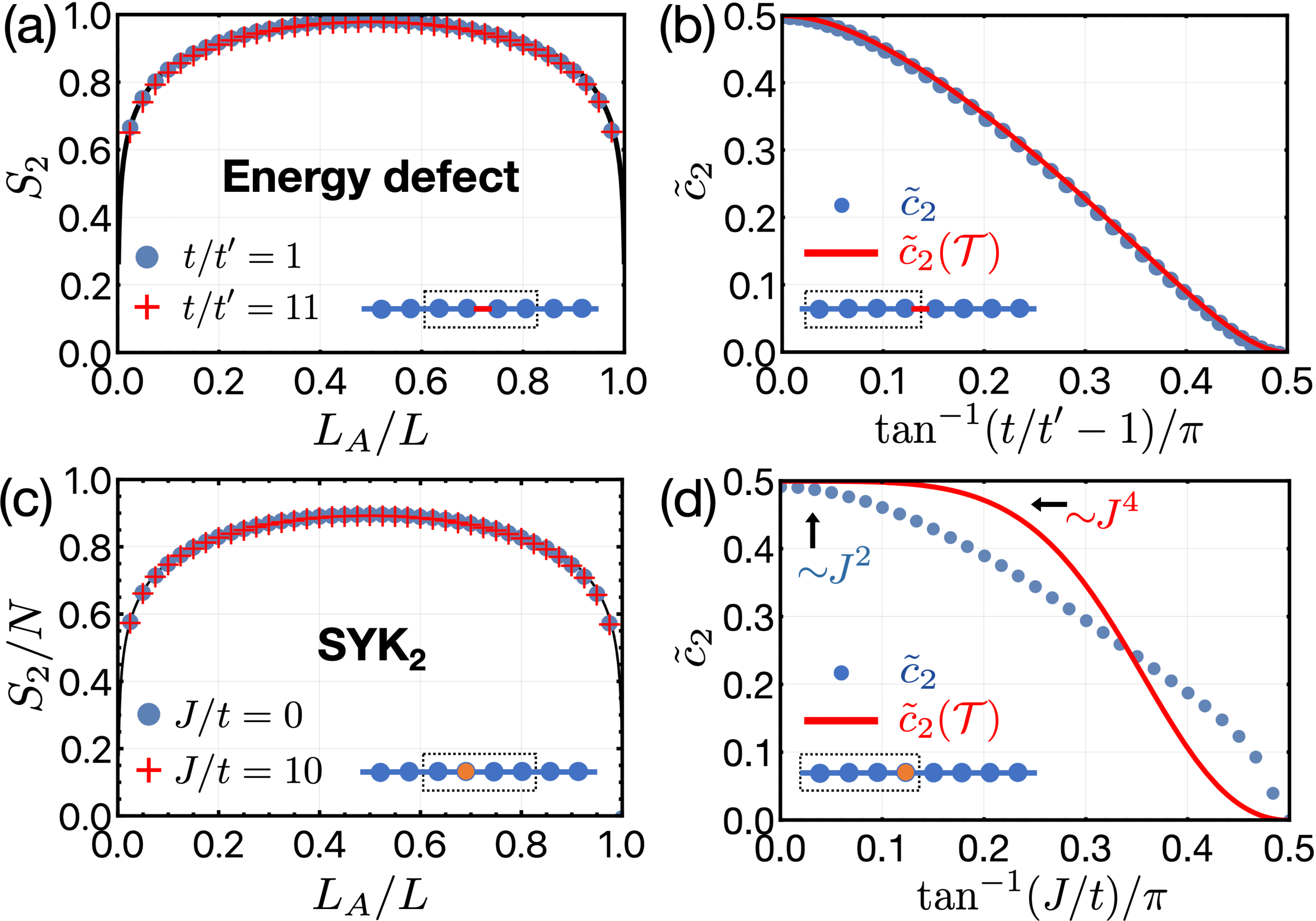}
    \caption{
    The second R\'enyi entropy and the extracted effective central charge under different partition schemes, in the presence of (a)--(b) an energy defect with bond strength $t'$, and (c)--(d) SYK$_2$ coupling of strength $J$ across $N$ Majorana chains, both at the center of the chain. 
    In (a) and (c), subsystem $A$ is the center region.
    Black curves give the analytical result of Eq.~\eqref{eq:S2frac} with $c_2=1/2$ and $L=4000$.
    In (b) and (d), the partition is at the defect. The effective central charge is computed both from the logarithmic scaling of $S_2$ (dots) and from the transmission probability $\mathcal{T}$ (line). Their disagreement in (d) signifies non-Gaussianity.
    }
    \label{fig:c2}
\end{figure}

\paragraph{Energy defect}\ssll
Two distinct bipartitions for the R\'enyi entropy that reveal universal conformal data have been considered in the literature: i) where the defect is located deep inside one subsystem, say $A$; ii) where the bipartition between two subsystems $A$ and $B$ is located at or near the defect. 
The first bipartition reveals the $g$ function via a folding trick, i.e., the offset of R\'enyi entropy between the system with and without defect equals the boundary entropy of a folded system with doubled degrees of freedom \cite{oshikawa1997boundary,saleur2002lectures,gutperle2016note}.
In the second bipartition, the entanglement  across the interface $S_2 \!\sim\! \frac{\tilde c_2}{8} \log L_A$, where $L_A$ is the size of region $A$ (for simplicity we set $L_B=L_A$). The prefactor $\tilde c_2$ defines the effective central charge, which can differ from the central charge of the bulk CFT.
In a free fermion CFT, the effective central charge is a universal function of the transmission coefficient $\mathcal T$~\cite{peschel2012exact}, i.e.,
\begin{equation}
    \tilde{c}_{2} = \frac{8}{\pi^2} \arcsin^2{\sqrt{\frac{\mathcal{T}}{2}}}.
    \label{eq:c2T}
\end{equation}
We shall see that this relation breaks down for an SYK defect.

Before returning to the SYK defect, we apply our method to evaluate the $g$ function and the effective central charge for the energy defect as a warmup~\cite{rogerson2022entanglement,oshikawa1996defect,oshikawa1997boundary}. 
The energy defect in the Ising CFT can be modeled by replacing the SYK site by a bond defect with a distinct hopping amplitude $t'$~\cite{peschel2012exact,rogerson2022entanglement}.
Let us set $N \! =\! 1$ since different chains do not couple in this case. 
With a slight modification, our saddle-point method can be applied to evaluate its R\'enyi entropy.
The result is shown in Fig.~\ref{fig:c2}(a)(b): 
Fig.~\ref{fig:c2}(a) shows that $\log g = 0$ for the energy defect, and Fig.~\ref{fig:c2}(b) shows the effective central charge, plotted along with the exact analytical result~\eqref{eq:c2T}, with $\sqrt{\mathcal T} = \frac2{t'/t + t/t'}$.
This benchmarks our method.

\paragraph{$g$ function}\ssll
To get the $g$ function in the presence of the SYK$_2$ defect, we put it in the middle of region $A$ in a symmetric setup, i.e., $L_1\!=\!L_2$ and $K_1\!=\!K_2$ in Fig.~\ref{fig:renyi-contour}(a), and compute $S_2$ as the subsystem size fraction varies while the total length is fixed. Since the $g$ function is universal, $S_2$ should depend logarithmically on $L_A$, the size of region $A$, with a constant offset when $\beta \!\gg\! L$ \cite{calabrese2004entanglement,rogerson2022entanglement},
\begin{equation}
    \frac{S_2}{N} = \frac{c_2}{4} \log \left[\frac{L}{\pi} \sin \left(\frac{\pi L_A}{L}\right) \right] + s,
    \label{eq:S2frac}
\end{equation}
such that the constant $s = \log g$ is independent of the defect strength $J$.
Here, the factor $c_2/4$ is the sum of the equal contribution $c_2/8$ from each of the two interfaces, and $N$ comes from the large-$N$ structure of the SYK.
The universality of $s$ is confirmed in Fig.~\ref{fig:c2}(c), which shows no offset between the entropy curves at different $J$'s
\footnote{A $J$-independent offset is added to the analytical curve in Fig.~\ref{fig:c2}(c) to match the result at $J \! = \! 0$. This offset comes from the imperfect mirror symmetry in even-length Majorana chains coupled at the SYK site, as seen in the insets of Fig.~\ref{fig:c2}. In contrast, no offset to the logarithm is needed for the energy defect in Fig.~\ref{fig:c2}(a).}, up to even-odd effects~\cite{fagotti2011parity,laflorencie2006boundary,affleck2009entanglement}.
This implies that $g \! = \! 1$, same as that of an energy defect \cite{delfino1994scattering}.

\paragraph{Effective central charge}\ssll
For the effective central charge due to the SYK defect, we partition our system at the SYK site as depicted in Fig.~\ref{fig:renyi-contour}(c). The result is a dCFT determined by the bulk free CFTs and the SYK defect \cite{eisler2010solution}. 
The second R\'enyi entanglement entropy across the SYK defect is $\frac{S_2}{N} \!\sim\! \frac{\tilde c_2}{8} \log L_A$~\cite{sakai2008entanglement,peschel2012exact,brehm2015entanglement,capizzi2022renyi}. %
Thus, we enlarge the system symmetrically in $A$ and $B$ to extract $\tilde{c}_2$. 
As seen in Fig.~\ref{fig:c2}(d), $\tilde c_2$ decreases towards zero at stronger $J$ until the two sides decouple. 
Therefore, the marginal SYK$_2$ interaction induces an interface CFT with a continuously tunable effective central charge. 
For irrelevant cases of $q\!>\!2$, the $\tilde{c}_2$ remains at $1/2$, that of a free Majorana chain or Ising CFT. 
A relevant coupling reduces $\tilde{c}_2$ to zero at low temperature, as is the case of massless bosonic chains with an SYK~\cite{Note1} or mass defect~\cite{lipkin1973quantum}, although the former is unstable~\cite{murugan2017more,liu2019d,tulipman2020strongly,shen2023long}.
Finite-$N$ calculations confirm the $q=2$ result qualitatively.

In contrast to free dCFTs, the transmission coefficient and the effective central charge are not related by \eqref{eq:c2T} for the SYK$_2$ defect, which is clear from Fig.~\ref{fig:c2}(d). The disagreement signifies the deviation from Gaussian defects due to the disorder averaging of SYK$_2$, even though each random realization is noninteracting.

\paragraph{Concluding remarks}\ssll
We have presented a novel family of boundary and defect CFTs, built from the SYK interaction coupling 1+1D systems, that exhibits a tunable effective central charge. 
Based on path integral and functional determinant, we devised a versatile method to compute the (conformal) data of defects embedded in 1+1D.
Our method admits arbitrary partitions of the system, and scales economically to low temperature and large systems. 
It can be extended to models with different boundary conditions, next-nearest-neighbor hoppings, etc. 
With this method, we evaluated the boundary entropy and effective central charge of the dCFT built from the SYK interaction.
Our findings suggest that it extends beyond all known Gaussian dCFTs.

While our primary focus is on the defect of the Ising CFT or free Majorana chain, it is worth emphasizing that our construction can be readily generalized to other large-$N$ models, including interacting ones. 
One example is the bosonic SYK defect, where a relevant defect coupling renders total reflection.
Another is the Yukawa-SYK defect~\cite{wang2020ysyk,ge2024ysyk}.
More importantly, nontrivial self-energy can be further incorporated to study cases with interacting bulk CFTs~\cite{Note1}. 
Beyond SYK-like models, our construction also opens an interesting arena to realize new defect states in CFTs such as the minimal models~\cite{jeng2001random}. 
For instance, random couplings between primary fields from $N$ identical copies of a minimal model could lead to a nontrivial renormalization group flow. 
We leave a detailed study for future work.

\begin{acknowledgments}
\paragraph{Acknowledgements}\ssll
We thank Xiao-Yang Shen for helpful discussions on the bosonic SYK model. 
This work is supported by a startup grant from Tulane University.
\end{acknowledgments}

\bibliography{references.bib}

\onecolumngrid

\setcounter{secnumdepth}{3}
\setcounter{equation}{0}
\setcounter{figure}{0}
\renewcommand{\theequation}{S\arabic{equation}}
\renewcommand{\thefigure}{S\arabic{figure}}
\renewcommand\figurename{Supplementary Figure}
\renewcommand\tablename{Supplementary Table}

\section*{Supplemental Material}

\section{Derivation of the large-$N$ action \label{append:large-N}} 

The model is given by the Hamiltonian $H=H_\text{CFT} + H_\text{I}$, where $H_\text{CFT}$ consists of $N$ decoupled Majorana chains, 
\begin{eqnarray}
	H_\text{CFT} =  -i 2 t \sum_r \psi_{j,r} \psi_{j,r+1}, \label{eq:ss:Hcft}
\end{eqnarray}
and $H_\text{I}$ denotes the interface given by SYK interaction
\begin{eqnarray}
	H_\text{I} = i^{q/2} \sum_{j_1,..., j_q}  J_{j_1,...,j_q} \psi_{j_1,0} ...\psi_{j_q,0}.
\end{eqnarray}
Here, $r$ is the site index, and $j=1,...,N$ denotes the flavor of Majorana at each site. Namely, each site has $N$ Majorana fermions. 
$\psi_{j,r}$ denotes the $j$-th Majorana fermion at site $r$,
and $\{\psi_{j,r}, \psi_{j',r'}\} = \delta_{j,j'} \delta_{r,r'}$. $t$ is the hopping amplitude between nearest-neighbor sites.
The SYK interaction that couples different chains exists at the site $r=0$. 
We refer to it as the SYK site. 

The interaction strength $J_{j_1,...,j_q}$ is a Gaussian variable with mean zero and variance
\begin{eqnarray} \label{seq:variance}
	\overline{J_{j_1,...,j_q} J_{j'_1,...,j'_q}} = \delta_{j_1, j'_1} ... \delta_{j_q, j'_q} \frac{2^{q-1} J^2 (q-1)!}{N^{q-1}}.
\end{eqnarray}

Define the hopping matrix  $h_{r_1, r_2} = \delta_{r_2, r_1 + 1} - \delta_{r_2, r_1 - 1}$. The action is thus
\begin{equation}
	- I =  - \frac12 \sum_{j, r_1, r_2} \int d\tau \, \psi_{j,r_1}(\tau)  (\partial_{\tau}  \delta_{r_1, r_2} - i t h_{r_1, r_2}) \psi_{j,r_2}(\tau) + \int d\tau \, i^{q/2} \sum_{j_1,..., j_q}  J_{j_1,...,j_q} \psi_{j_1,0} ...\psi_{j_q,0}.
\end{equation}
After integrating over the Gaussian distributed interaction, the action reads
\begin{equation}
	- I =  - \frac12 \sum_{j, r_1, r_2} \int d\tau \, \psi_{j,r_1}(\tau)  (\partial_{\tau}  \delta_{r_1, r_2} - i t h_{r_1, r_2}) \psi_{j,r_2}(\tau) + \frac{N J^2}{4q} \int d\tau_1 d \tau_2 \left(2 \frac1{N} \sum_j \psi_{j,0}(\tau_1) \psi_{j,0}(\tau_2) \right)^q,
\end{equation}
To derive the $(G,\Sigma)$ action, we introduce the bilocal fields $G(\tau_1, \tau_1)$, $\Sigma (\tau_1, \tau_2)$ and multiply the partition function by a constant \cite{maldacena2016remarks,bagrets2016SYK,gu2017spread,kitaev2018soft},
\begin{equation}
	1=\int \mathcal{D}G \mathcal{D}\Sigma \, e^{-\frac{N}{2}\int d\tau_1 d \tau_2 \, G(\tau_1,\tau_2)\Sigma(\tau_1,\tau_2)} = \int \mathcal{D}G \, \delta[G(\tau_1,\tau_2)].
\end{equation}
Here the measure $\mathcal{D}G \mathcal{D}\Sigma \sim \prod_{\tau_1,\tau_2}\frac{N}{2\pi}\mathrm{d}G(\tau_1,\tau_2)\mathrm{d}[i\Sigma(\tau_1,\tau_2)]$.
Then we shift $G(\tau_1,\tau_2) \to G(\tau_1,\tau_2)-\sum_j \psi_{j,0}(\tau_1) \psi_{j,0}(\tau_2)/N$. By virtual of the delta function, $G$ plays the role of the Majorana propagator. This also shifts the quadratic Hamiltonian by $\Sigma$, which one can later identify with the self-energy in the Schwinger-Dyson equation \eqref{eq:gandsig}. The result is
\begin{eqnarray}
	- I &=&  - \frac12 \sum_{j, r_1, r_2} \int d\tau_1 d \tau_2 \, \psi_{j,r_1}(\tau_1)  \left[(\partial_{\tau_1}  \delta_{r_1, r_2} - i t h_{r_1, r_2} ) \delta(\tau_1 - \tau_2) - \delta_{r_1, 0} \delta_{r_2, 0}\Sigma(\tau_1, \tau_2) \right] \psi_{j,r_2}(\tau_2) \nonumber \\
	&& - \frac{N}2 \int d\tau_1 d\tau_2 \, G(\tau_1, \tau_2) \Sigma(\tau_1, \tau_2) + \frac{N J^2}{4q} \int d\tau_1 d \tau_2 \left[2 G(\tau_1, \tau_2) \right]^q.
\end{eqnarray}
It is not hard to check that, by integrating over $G$ and $\Sigma$, this reduces to the fermionic action.
Now the action is only quadratic in terms of Majorana fermions, so we can integrate them out to get
\begin{eqnarray}
	- \frac{I}N &=&  \frac12  \log \det \left[(\partial_{\tau_1}  \delta_{r_1, r_2} - i t h_{r_1, r_2} ) \delta(\tau_1 - \tau_2) - \delta_{r_1, 0} \delta_{r_2, 0}\Sigma(\tau_1, \tau_2) \right]  \nonumber \\
	&& - \frac{1}2 \int d\tau_1 d\tau_2 \, G(\tau_1, \tau_2) \Sigma(\tau_1, \tau_2) + \frac{J^2}{4q} \int d\tau_1 d \tau_2 \left[2 G(\tau_1, \tau_2) \right]^q.
\end{eqnarray}
The determinant involves matrices whose indices range over the imaginary time and the lattice sites.
Since the action has a large-$N$ structure, we can implement a saddle point analysis.
The Schwinger-Dyson equations read
\begin{eqnarray}
	G(\tau_1, \tau_2) = (\partial_\tau - \Sigma - i t h)^{-1}_{00}(\tau_1, \tau_2), \quad \Sigma(\tau_1, \tau_2) = J^2 [2G(\tau_1, \tau_2)]^{q-1}, \label{eq:gandsig}
\end{eqnarray}
where the first equation should be understood as a matrix equation.
The subscript denotes $r_1 = 0$, $r_2 = 0$ component.

We can further simplify the action by noting that the self-energy is nontrivial only at the SYK site.
We assume the solution to be time translationally symmetric, i.e., $G(\tau_1, \tau_2) = G(\tau_1 - \tau_2)$, $\Sigma(\tau_1, \tau_2) = \Sigma(\tau_1 - \tau_2)$.
In this case, we can perform a Fourier transformation in the imaginary time domain, i.e.,
\begin{eqnarray}
	G(\tau_1, \tau_2) = \frac1{\beta} \sum_n e^{i \omega_n (\tau_1 - \tau_2)} G(i\omega_n), \quad \Sigma(\tau_1, \tau_2) = \frac1{\beta} \sum_n e^{i \omega_n (\tau_1 - \tau_2)} \Sigma(i\omega_n), \quad \omega_n = \frac{(2n+1)\pi}{\beta}.  
\end{eqnarray}
We consider the open boundary condition for these Majorana chains. In the frequency space, the matrix inside the determinant becomes tridiagonal in the basis of $\psi_r(i\omega_n)$ at each Matsubara frequency,
\begin{equation}
	\mathcal{M}_{rr'} = \bigoplus_n \left[ -i\omega_n \delta_{rr'} - \Sigma(i\omega_n) \delta_{r,0}\delta_{0,r'}- it \delta_{r+1,r'} + it \delta_{r,r'+1} \right].
\end{equation}
The determinant of a tridiagonal matrix has a simple analytical expression~\cite{mallik2001inverse}.
Therefore, we can first evaluate the determinant at each frequency, and then take the product over all the frequencies.
After a straightforward calculation, the large-$N$ action can be simplified as 
\begin{eqnarray}
	- \frac{I}N &=&  \frac12  \log \prod_n \left[ -i\omega_n - \Sigma(i\omega_n) - t R_{K}\left(\frac{-i\omega_n}{2t}\right)- t R_{L}\left(\frac{-i\omega_n}{2t}\right) \right]  \nonumber\\
	&& - \frac{1}2 \int d\tau_1 d\tau_2 \, G(\tau_1, \tau_2) \Sigma(\tau_1, \tau_2) + \frac{J^2}{4q} \int d\tau_1 d \tau_2 \left[2 G(\tau_1, \tau_2) \right]^q + \frac{I'}{N}.
	\label{eq:I}
\end{eqnarray}
where $R_L(x) = \frac{U_{L-1}(x)}{U_L(x)}$, and $U_L$ is the Chebyshev polynomial of the second kind.
$K-1$ and $L$ are the number of sites to the left and right of the SYK sites, respectively, as shown in Fig.~\ref{fig:model}. The term 
\begin{equation}
	\frac{I'}{N}=\frac{1}{2} \sum_n \log \left[ t^{L+K} U_L\left(\frac{-i\omega_n}{2t}\right) U_{K}\left(\frac{-i\omega_n}{2t}\right) \right]
\end{equation}
is independent of $G$ and $\Sigma$. Varying the bilocal fields, we obtain the corresponding Schwinger-Dyson equation
\begin{eqnarray}
	G(i\omega_n) &=& \left[ -i\omega_n - \Sigma(i\omega_n) - t R_{K}\left(\frac{-i\omega_n}{2t}\right)- t R_{L}\left(\frac{-i\omega_n}{2t} \right) \right]^{-1}, \label{eq:dyson:old} \\
	\Sigma(\tau_1, \tau_2) &=& J^2 [2G(\tau_1, \tau_2)]^{q-1}. \label{eq:Sigq}
\end{eqnarray}
The effect of coupling to the Majorana chain is reflected in the correction to the self-energy in the form of ratios of the Chebyshev polynomials. Finally, we note that in the numerics, the ratio of $U_L$'s with large $L$'s can be numerically unstable, and the recursive relation $R_L(x)=1/[2x-R_{L-1}(x)]$ may be helpful.

\section{Compute the Green's function at $L=\infty$ for any $q$ \label{append:saddle-point}}

Here we derive in detail the large-$N$ Green's function and self-energy of infinitely long SYK$_q$-coupled Majorana chains, which are Eqs.~\eqref{eq:dyson:old} and \eqref{eq:Sigq} or Eqs.~\eqref{eq:dyson} in the main text.
To proceed, we consider an infinite number of sites to the left and right of the SYK point, i.e., $L, K \rightarrow \infty$.
Observing that
\begin{eqnarray}
	\lim_{L \rightarrow \infty} R_L(x) = x - \sqrt{x^2 - 1}, 
\end{eqnarray}
the Schwinger-Dyson equation~\eqref{eq:dyson:old} becomes
\begin{eqnarray}
	G(i\omega_n) = \left[  \sqrt{- 4t^2 - \omega_n^2} - \Sigma(i\omega_n)  \right]^{-1}.
\end{eqnarray}
Note that $\sqrt{-4t^2-\omega_n^2}$ is simply the $G^{-1}(i\omega_n)$ of a 1D chain. Unlike the SYK model, the coupling to the Majorana chain is relevant at small frequencies for $q > 2$, while the SYK$_q$ type of interaction is irrelevant for $q > 2$.
Therefore, we consider SYK$_2$ where the interaction is marginal. 
To this end, we eliminate $\Sigma(i \omega_n)$ so that
\begin{eqnarray}
	G(i\omega_n) = \left[ \sqrt{-4t^2 - \omega_n^2} - 2 J^2 G(i \omega_n)  \right]^{-1}.
\end{eqnarray}
Solving it for the advanced Green's function gives
\begin{eqnarray}
	G(\omega-i\eta) = \frac{i}{4J^2} \left[ \sqrt{4t^2-(\omega-i\eta)^2}-\sqrt{4t^2+8J^2-(\omega-i\eta)^2}  \right].
	\label{eq:inf}
\end{eqnarray}
The associated spectral function is shown in Fig.~\ref{fig:infinite}(a).

For the interacting cases $q>2$, we expand $G$ using its spectral function $A(\omega)= 2G''(\omega-i\eta)$ in Eq.~\eqref{eq:inf}.
The result for the self-energy at any $q$ is
\begin{eqnarray}
	\label{seq:sigq-om}
	\Sigma''_q(\omega-i\eta) = J^2  \int \left[ \prod_{i = 1}^{q - 2} \frac{\mathrm{d} \nu_i}{\pi} 
	A (\nu_i) \right] A \left( \omega - \textstyle\sum_{j = 1}^{q - 2} \nu_j \right)  \prod_{k = 1}^{q - 2} \left[ n_{\zeta_k} \left( \textstyle\sum_{l = k}^{q - 2} \nu_l- \omega \right) - n_F (\nu_k)  \right],
\end{eqnarray}
where
\begin{eqnarray}
	n_{\zeta_i} (\nu) = \left\{\begin{aligned}%
		-n_B & (\nu), \  i \in 2\mathbb{Z},\\
		n_F & (\nu), \ i \in 2\mathbb{Z}+ 1.
	\end{aligned}\right.
\end{eqnarray}
Hilbert transform of the spectral function recovers the full Green's function.%
Numerical computations of the integrals can be formulated into convolutions that are expedited with the fast Fourier transform. For example, the self-energy of SYK$_4$ is given by multiplications and convolutions in real frequencies.
Let the convolution be denoted by $\ast$, and omit the frequency arguments of $A$, $n_F$, and $n_B$, as they are identical.
The expression for the self-energy is then given by
\begin{eqnarray}
	\Sigma''_4(\omega-i\eta)=\frac{J^2}{\pi^2}  \left[ (n_F A) \ast A \ast A + ( \{ [(1-2n_F) A] \ast A \} \{1+n_B\} ) \ast A \right].
\end{eqnarray}
The local spectral function at the SYK$_4$ site is shown in Fig.~\ref{fig:infinite}(b)(c). Note that at high temperature, the Majorana chain-coupled SYK$_4$ dot behaves similarly to a 0+1D SYK dot. At low temperature, it flows away from the SYK point.

\section{Derivation of the second R\'enyi entropy}
\label{append:renyi}

To compute the R\'enyi entropy, we first assume that the chains satisfy an open boundary condition. 
Then, we divide the system into two regions, A and B. We are interested in the entanglement entropy between these two regions. The computation of entanglement entropy employs the replica trick \cite{calabrese2004entanglement}. Below, we first review the replica trick and then derive the R\'enyi entanglement entropy for the 1+1D chains.

\subsection{Review of the replica trick}

\begin{figure}
	\includegraphics[width=0.12\linewidth]{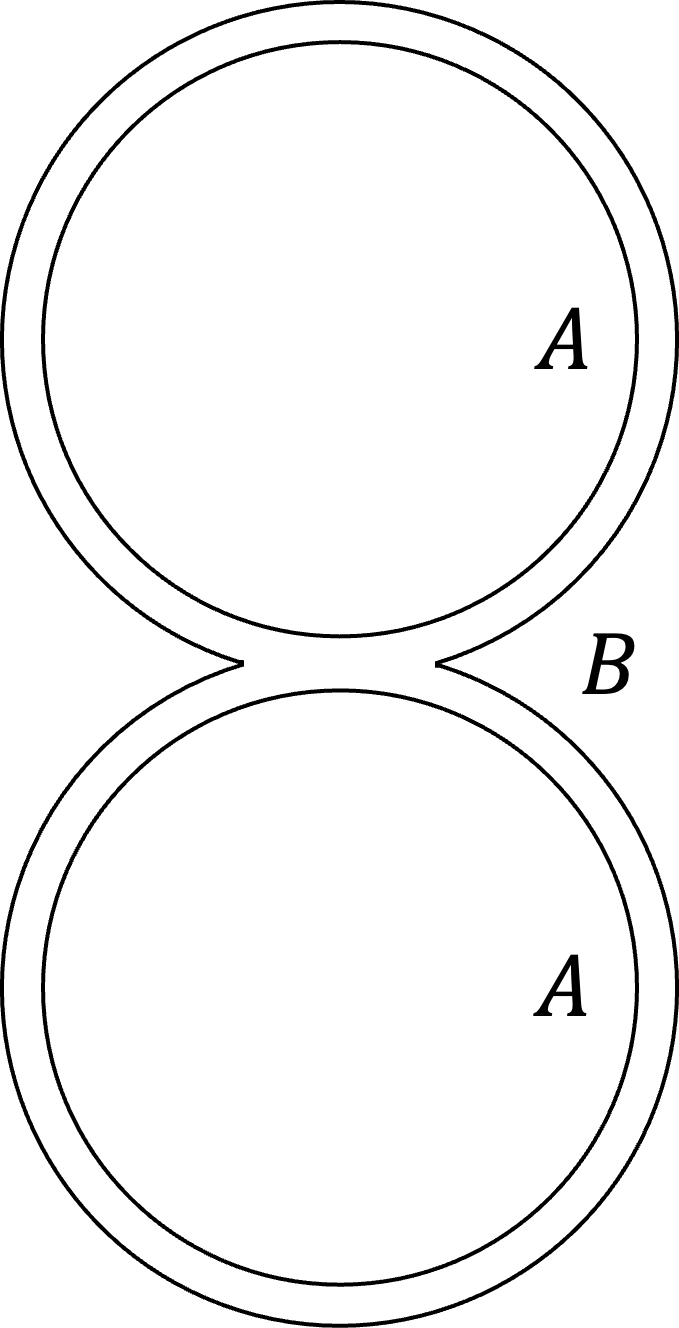}
	\caption{The imaginary time contour for the second R\'enyi entropy.
		The fermionic field in a closed contour satisfies the conventional anti-periodic boundary condition, that leads to the fermionic Matsubara frequency.}
	\label{fig:contour}
\end{figure}

We consider the thermal density matrix $\rho = \frac{e^{-\beta H}}{Z}$, $Z = \Tr(e^{-\beta H})$.
The second R\'enyi entropy of the region $A$ is
\begin{eqnarray}
	S_2 = - \log \Tr_A[ (\Tr_B \rho )^2 ] = -\log \frac{Z_{(2)}}{Z^2}.
\end{eqnarray}
Here, $Z_{(2)}$ denotes a path integral with a twist boundary condition on the region $A$.
In terms of replica fields, the boundary condition reads
\begin{alignat}{2}
	\psi^{(1)}_{j \in A}( \beta ) &= -\psi^{(1)}_{j \in A}( 0 ),& \quad 
	\psi^{(2)}_{j \in A}( \beta ) &= - \psi^{(2)}_{j \in A}( 0 ), \\
	\psi^{(1)}_{j \in B}( \beta ) &= \psi^{(2)}_{j \in B}( 0 ),& \quad 
	\psi^{(2)}_{j \in B}( \beta ) &= - \psi^{(1)}_{j \in B}( 0 ),
\end{alignat}
where the superscripts $1, 2$ denote the two replicas.
The imaginary time contour for $Z_{(2)}$ is shown in Fig.~\ref{fig:contour}. 
The fermionic field in a closed contour satisfies the conventional anti-periodic boundary condition that gives rise to the fermionic Matsubara frequency.
A crucial point is that the closed imaginary-time contours of the fermionic fields in regions $A$ and $B$ have different periods.
To this end, we introduce a different parametrization of the imaginary time contour, $s \in (0, 2\beta)$, such that $s<\beta$ ($s>\beta$) denotes the first (second) replica, i.e., 
\begin{eqnarray}
	\psi_j(s) = \begin{cases} \psi_j^{(1)}(s), & s \in (0, \beta), \\ 
		\psi_j^{(2)}(s-\beta), & s \in (\beta, 2\beta).
	\end{cases}
	\label{eq:field}
\end{eqnarray}
The boundary condition now becomes
\begin{eqnarray} 
	\psi_{j\in A}(\beta_-) = - \psi_{j\in A} (0_+), \quad \psi_{j\in A}(2\beta_-) = - \psi_{j\in A} (\beta_+), \quad \psi_{j\in B}(2\beta_-) = - \psi_{j\in A} (0_+).
\end{eqnarray}
The fermionic field in the region $A$ can be expanded using the fermionic mode with the Matsubara frequency $\Omega_n = \frac{(2n+1)\pi}{2\beta}$, while the fermionic field in the region $B$ needs to be expanded by two fermionic modes with the Matsubara frequency $\omega_n = \frac{(2n+1)\pi}{\beta}$. 
More explicitly, we have the following expansions,
\begin{eqnarray}
	\psi_{j\in A}(s) & = & \frac1{\sqrt{\beta}} \sum_n \left[ \psi_{j\in A}^{(1)}(i\omega_n)  e^{-i\omega_n s} \Theta(\beta - s) + \psi_{j\in A}^{(2)}(i\omega_n)  e^{-i\omega_n s} \Theta(s - \beta) \right], \quad \omega_n = \frac{(2n+1)\pi}{\beta},\\
	\psi_{j\in B}(s) & = & \frac1{\sqrt{2\beta}} \sum_n \psi_{j\in B}(i\Omega_n)  e^{-i\Omega_n s}, \quad \Omega_n = \frac{(2n+1)\pi}{2\beta}.
\end{eqnarray}
At the interface of regions $A$ and $B$, the fields on the two sides are glued together by the twist operator $\sigma$, defined by
\begin{equation}
	\sigma (i \omega_m^{(a)}, i \Omega_n) = 
	\int^{a\beta}_{(a-1)\beta} \frac{\mathd \tau}{\sqrt{2} \beta} e^{i [\omega_m^{(a)}
		- \Omega_n] \tau}, \quad a=1,2,
	\label{eq:twist}
\end{equation}
so that $\sigma (i \omega_m^{(a)}, i \Omega_n)$ is the overlap between the modes $\psi_{j\in A}^{(a)}(i\omega_m)$ and $\psi_{j\in B}(i\Omega_n)$.
To prepare for the derivation of the replica action and the corresponding Green's functions, we also define the diagonal matrices of Matsubara frequencies $\boldsymbol{\omega}\coloneq\left(\bigoplus_n\omega_n\right)\otimes\mathbf{1}_2$, $\boldsymbol{\Omega}\coloneq\bigoplus_n\Omega_n$.

\subsection{Saddle-point solution to the replica action}\label{ss:island}

\begin{figure}[h!]
	\centering
	\includegraphics[width=0.5\linewidth]{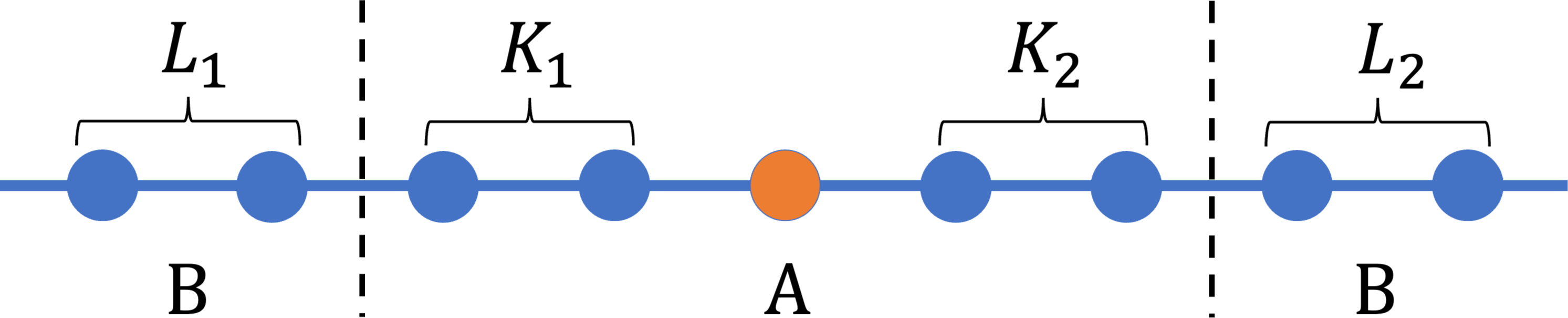}
	\caption{Geometry for the SYK island. The system is partitioned into regions $A$ and $B$ for all of the $N$ chains. All the chains are coupled at the SYK site, marked by orange. For simplicity only one chain is drawn. }
	\label{fig:island}
\end{figure}

Consider the generic partition on the 1+1D chain, as depicted in Fig.~\ref{fig:island} [Fig.~\ref{fig:renyi-contour}(a) in main text]. We compute the entanglement entropy between regions $A$ and $B$.

We again introduce the bilocal fields $G_{(2)}$ and $\Sigma_{(2)}$ at the SYK site, and then integrate out the free Majorana fields on the rest of the chain. The action becomes
\begin{equation}
	- \frac{I_{(2)}}N =  \frac12  \log \det \mathcal{M} - \frac{1}2 \int ds_1 ds_2 \, G_{(2)}(s_1, s_2) \Sigma_{(2)}(s_1, s_2) + \frac{J^2}{4q} \int ds_1 d s_2 \left[2 G_{(2)}(s_1, s_2) \right]^q,
\end{equation}
where $\mathcal{M}$ is given by
\begin{equation}
	\mathcal{M}=
	\left( \begin{array}{cr|lcccr|lc}
		\multicolumn{2}{c|}{\multirow{2}{*}{\large{\ensuremath{B_1}}}} & & & & & & & \\
		& & \ensuremath{- i r_1 t \sigma^{\dagger}} & & & & & &   \\
		\hline
		\ensuremath{\ \ } & \ensuremath{i r_1 t \sigma} & \multicolumn{2}{c|}{\multirow{2}{*}{\large{\ensuremath{A_1}}}}  & & & & &  \\ 
		& & & \multicolumn{1}{c|}{} & \multicolumn{1}{c}{\ensuremath{- i q_1 t}}  & & & \\ \cline{3-4}
		& & &  \multicolumn{1}{c}{\ensuremath{i q_1 t}} & \ensuremath{-i\boldsymbol{\omega} \!-\! \Sigma} & \ensuremath{- i q_2 t} & & &\\ \cline{6-7}
		& & & & \ensuremath{i q_2 t} & \multicolumn{2}{|c|}{\multirow{2}{*}{\large{\ensuremath{A_2}}}} & & \\
		& & & & & \multicolumn{1}{|c}{} & & \ensuremath{- i r_2 t \sigma^{\dag}} & \ensuremath{\ \ } \\
		\hline
		& & & & & & \ensuremath{i r_2 t \sigma} & \multicolumn{2}{c}{\multirow{2}{*}{\large{\ensuremath{B_2}}}} \\
		& & & & & & &
	\end{array} \right).
\end{equation}
To make the result more versatile, we parameterize the hopping strengths between regions $A$ and $B$ by $r_{1,2}$ at the left and right interfaces, respectively. The hoppings to the SYK site from the left and right are also parameterized by $q_{1,2}$, respectively. This extends our method to, e.g., the energy defect.

Discretizing imaginary time into an even number of points $M$, the sizes of the blocks $A_{1, 2}$ are $K_{1, 2} M \times K_{1, 2} M$.
Similarly, the sizes of $B_{1, 2}$ are $L_{1, 2} M \times L_{1, 2} M$,
and the sizes of $-i\boldsymbol{\omega} - \Sigma$, $\pm it$, and $\pm it \sigma$ are
all $M \times M$. The $\pm it$ blocks are proportional to identity in frequency space. The center
block represents the region $A$.

The determinant can be factored using $\det \begin{psmallmatrix}
	A & B\\
	C & D
\end{psmallmatrix}=\det A \det (D-C A^{-1} B)$ when $A$ and $D$ are square matrices. Denote the special matrices $(e_n)_{ij}:=\delta_{i,n}\delta_{n,j}$. Then,
\begin{eqnarray}
	\det \mathcal{M} & = & \det B_1 \det B_2 \det \left[ A - r^2_1 t \sigma
	R_{L_1} \left( \tfrac{- i\boldsymbol{\Omega}}{2 t} \right) \sigma^{\dagger} e_1 -
	r^2_2 t \sigma R_{L_2} \left( \tfrac{- i\boldsymbol{\Omega}}{2 t} \right)
	\sigma^{\dagger} e_{K_1 + K_2 + 1} \right] \nonumber\\
	& = & \det B_1 \det B_2 \det A_1' \det A_2' \det \left[ -i\boldsymbol{\omega} - \Sigma_{(2)} -
	q_1^2 t^2  \left( {A'_1}^{- 1} \right)_{K_1, K_1} - q_2^2 t^2  \left( {A'_2}^{- 1}
	\right)_{1, 1}  \right] . 
\end{eqnarray}
Here we denoted $A'_1 = A_1 - r^2_1 t \sigma A_{L_{1}} \left(
\frac{- i\boldsymbol{\Omega}}{2 t} \right) \sigma^{\dagger} e_1$, $A'_2 = A_2 - r^2_2 t \sigma R_{L_{2}} \left(
\frac{- i\boldsymbol{\Omega}}{2 t} \right) \sigma^{\dagger} e_{K_2}$. The
diagonal corner entries of $A'^{-1}$'s can be obtained from the formula \cite{zhang2013matrix}
\begin{equation}
	\begin{bmatrix}
		A & B\\
		C & D
	\end{bmatrix}^{- 1} = \begin{bmatrix}
		(A - BD^{- 1} C)^{- 1} & 0 \\
		0 & (D - CA^{- 1} B)^{- 1}
	\end{bmatrix}
	\begin{bmatrix}
		\mathbf{1} & -BD^{-1} \\
		-CA^{-1} & \mathbf{1}
	\end{bmatrix}.
\end{equation}
This gives
\begin{eqnarray}
	\left( {A'_1}^{- 1} \right)_{K_1, K_1} & = & \frac{1}{t}  \cfrac{1}{-
		\frac{i\boldsymbol{\omega}}{t} - \cfrac{1}{- \frac{i\boldsymbol{\omega}}{t} -
			\cfrac{1}{\cdots - \cfrac{1}{- \frac{i\boldsymbol{\omega}}{t} - r^2_1 \sigma R_{L_1}
					\left( \frac{- i\boldsymbol{\Omega}}{2 t} \right) \sigma^{\dagger}}}}} . 
\end{eqnarray}
In other words, let
\begin{eqnarray}
	\mathbf{D}_{0,L} & = & r^2 \sigma R_{L} \left( \frac{- i\boldsymbol{\Omega}}{2 t} \right)
	\sigma^{\dagger}, \nonumber\\
	\mathbf{D}_{K,L} & = & \left( - \frac{i\boldsymbol{\omega}}{t} - \mathbf{D}_{K - 1,L} \right)^{- 1} . 
\end{eqnarray}
Note that when $\mathbf{D}_{0,L} = 0$, $\mathbf{D}_{K,L} = R_K \left( \frac{- i\boldsymbol{\omega}}{2 t}
\right)$. The general formula is,
\begin{equation}
	\mathbf{D}_{K,L} = \left[ R_{K}\left( \frac{-i\boldsymbol{\omega}}{2t} \right)-U_{K-2}\left( \frac{-i\boldsymbol{\omega}}{2t} \right) \mathbf{D}_{0,L} U^{-1}_{K}\left( \frac{-i\boldsymbol{\omega}}{2t} \right)  \right]\left[1-U_{K-1}\left( \frac{-i\boldsymbol{\omega}}{2t} \right) \mathbf{D}_{0,L} U^{-1}_{K}\left( \frac{-i\boldsymbol{\omega}}{2t} \right) \right]^{-1}.
	\label{eq:ss:Dgeneral}
\end{equation}
Thus, $(A'_1)^{- 1} = \mathbf{D}_{K_1,L_1} / t$. Similar results hold for $(A_2')^{- 1}$. Furthermore,
\begin{eqnarray}
	\det B_1 & = & t^{L_1} U_{L_1} \left( \tfrac{- i\boldsymbol{\Omega}}{2 t}
	\right), \\
	\det A_1' & = & \det \left[ A_1 - r^2_1 t \sigma R_{L_1} \left( \tfrac{-
		i\boldsymbol{\Omega}}{2 t} \right) \sigma^{\dagger}  e_{1} \right] \nonumber\\
	& = & t^{K_1 - 1} U_{K_1 - 1} \left( \tfrac{- i\boldsymbol{\omega}}{2 t}
	\right) \det \left[ - i\boldsymbol{\omega}- tR_{K_1 - 1} \left( \tfrac{-
		i\boldsymbol{\omega}}{2 t} \right) - r^2_1 t \sigma R_{L_1} \left( \tfrac{-
		i\boldsymbol{\Omega}}{2 t} \right) \sigma^{\dagger}  \right] \nonumber\\
	& = & t^{K_1 - 1} U_{K_1 - 1} \left( \tfrac{- i\boldsymbol{\omega}}{2 t}
	\right) \det \left[ tR_{K_1}^{- 1} \left( \tfrac{- i\boldsymbol{\omega}}{2 t}
	\right) - r^2_1 t \sigma R_{L_1} \left( \tfrac{- i\boldsymbol{\Omega}}{2 t}
	\right) \sigma^{\dagger}  \right] .
\end{eqnarray}
Therefore,
\begin{eqnarray}
	- \frac{I_{(2)}}{N} & = & \frac{1}{2} \log \det [- i\boldsymbol{\omega} - \Sigma_{(2)} - q^2_1 t\mathbf{D}_{K_1,L_1}
	- q^2_2  t\mathbf{D}_{K_2,L_2} ] \nonumber\\
	&  & - \frac{1}{2}  \int d \tau_1 d \tau_2  \hspace{0.17em} G (\tau_1,
	\tau_2) \Sigma_{(2)} (\tau_1, \tau_2) + \frac{J^2}{4 q}  \int d \tau_1 d \tau_2 
	[2 G (\tau_1, \tau_2)]^q \nonumber\\
	&  & + \frac{1}{2} \log \det \left[ tR_{K_1}^{- 1} \left( \tfrac{-
		i\boldsymbol{\omega}}{2 t} \right) - r^2_1 t \sigma R_{L_1} \left( \tfrac{-
		i\boldsymbol{\Omega}}{2 t} \right) \sigma^{\dagger} \right] + \frac{1}{2} \log
	\det \left[ tR_{K_2}^{- 1} \left( \tfrac{- i\boldsymbol{\omega}}{2 t} \right) -
	r^2_2 t \sigma R_{L_2} \left( \tfrac{- i\boldsymbol{\Omega}}{2 t} \right)
	\sigma^{\dagger} \right] \nonumber\\
	&  & + \frac{1}{2} \log \det \left[ t^{L_1 + L_2 + K_1 + K_2 - 2} U_{L_1}
	\left( \tfrac{- i\boldsymbol{\Omega}}{2 t} \right) U_{L_2} \left( \tfrac{-
		i\boldsymbol{\Omega}}{2 t} \right) U_{K_1 - 1} \left( \tfrac{-
		i\boldsymbol{\omega}}{2 t} \right) U_{K_2 - 1} \left( \tfrac{-
		i\boldsymbol{\omega}}{2 t} \right) \right] .
	\label{eq:islandI2}
\end{eqnarray}
The saddle point equations is
\begin{eqnarray}
	G_{(2)}^{- 1} & = & - i\boldsymbol{\omega}- \Sigma_{(2)} - q^2_1 t\mathbf{D}_{K_1,L_1} - q^2_2 t\mathbf{D}_{K_2,L_2} . 
\end{eqnarray}
On the other hand, the untwisted action on a single replica can be obtained by modifying Eq.~\eqref{eq:I}, 
\begin{eqnarray}
	- \frac{I}{N} & = & \frac{1}{2} \sum_n \log \left[- i{\omega_n} - \Sigma(\omega_n) - q_1^2 t {D}'_{K_1,L_1}\left(\tfrac{-i\omega_n}{2t}\right) -
	q_2^2 t {D}'_{K_2,L_2}\left(\tfrac{-i\omega_n}{2t}\right) \right] \nonumber\\
	&  & - \frac{1}{2}  \int d \tau_1 d \tau_2  \hspace{0.17em} G (\tau_1,
	\tau_2) \Sigma (\tau_1, \tau_2) + \frac{J^2}{4 q}  \int d \tau_1 d \tau_2 
	[2 G (\tau_1, \tau_2)]^q \nonumber\\
	&  & + \frac{1}{2} \sum_n \log \left[ tR_{K_1}^{- 1} \left(\tfrac{-i\omega_n}{2t}\right) - r^2_1 t R_{L_1} \left(\tfrac{-i\omega_n}{2t}\right) \right] + \frac{1}{2} \sum_n \log \left[
	tR_{K_2}^{- 1} \left(\tfrac{-i\omega_n}{2t}\right) - r^2_2
	tR_{L_2} \left(\tfrac{-i\omega_n}{2t}\right) \right] \nonumber\\
	&  & + \frac{1}{2} \sum_n \log \left[ t^{L_1 + L_2 + K_1 + K_2 - 2} U_{L_1}
	\left(\tfrac{-i\omega_n}{2t}\right) U_{L_2} \left(\tfrac{-i\omega_n}{2t}\right) U_{K_1 - 1} \left(\tfrac{-i\omega_n}{2t}\right) U_{K_2 - 1} \left(\tfrac{-i\omega_n}{2t}\right) \right],
	\label{eq:islandI0}
\end{eqnarray}
where ${D}'_{K,L}$ has the same recursive structure as $\mathbf{D}_{K,L}$, with ${D}'_{0,L}\left(\frac{-i\omega_n}{2t}\right) = r^2R_{L}\left(\frac{-i\omega_n}{2t}\right)$. Since all terms commute, it can be simplified into
\begin{equation}
	{D}'_{K,L}\left(\frac{-i\omega_n}{2t}\right) \coloneq \frac{1-R_{K-1}\left( \frac{-i\boldsymbol{\omega}}{2t} \right) r^2 R_{L}\left( \frac{-i\boldsymbol{\omega}}{2t} \right) }{R^{-1}_{K}\left( \frac{-i\boldsymbol{\omega}}{2t} \right)-r^2 R_{L}\left( \frac{-i\boldsymbol{\omega}}{2t} \right) }.
\end{equation}

The second R\'enyi entropy is given by $S_2 = I_{(2)} - 2I$. We can further simplify the expression for the ground-state R\'enyi entropy by subtracting the zero $S_2$ when A and B are decoupled, thus dropping the last terms in \eqref{eq:islandI2} and \eqref{eq:islandI0}. At $r = 0$, $\mathbf{D}_{K} = {D}_{K}'\left(\frac{-i\boldsymbol{\omega}}{2t}\right) = R_{K}
\left( \frac{- i\boldsymbol{\omega}}{2 t} \right)$. Denote the entropy terms in
a decoupled system by $\check{I}$, and also let $\tilde{G}$ be the Green's function $G$ in \eqref{eq:dyson} repeated on each replica, $\tilde{G}=G\otimes\mathbf{1}_2$, such that
\begin{equation}
	\tilde{G}(\tau_1,\tau_2)=\begin{cases}
		G(\tau_1,\tau_2), & 0 \leq \tau_1, \tau_2 \leq \beta, \\ 
		G(\tau_1-\beta,\tau_2-\beta), & \beta \leq \tau_1, \tau_2 \leq 2\beta, \\ 
		0, & \text{otherwise}.
	\end{cases}
\end{equation}
Then, we arrive at the ground-state R\'enyi entropy
\begin{eqnarray}
	S_2 / N & = & \left[ I_{(2)} - 2I - (\check{I}_{(2)} - 2\check{I})\right] / N \nonumber\\
	& = & - \frac{1}{2} \log \det [G_{(2)}^{- 1} \tilde{G}] + J^2  \left( \frac{1}{4 q} - \frac{1}{4} \right) \int \mathd \tau_1 \mathd \tau_2  \{ [2 \tilde{G} (\tau_1, \tau_2)]^q - [2 G_{(2)} (\tau_1, \tau_2)]^q \} \nonumber\\
	&  & + \frac{1}{2} \log \det \left[ \frac{1 - r^2_1 R_{L_1} \left( \frac{- i\boldsymbol{\omega}}{2 t} \right) R_{K_1} \left( \frac{- i\boldsymbol{\omega}}{2 t} \right)}{1 - r^2_1 \sigma R_{L_1} \left( \frac{- i\boldsymbol{\Omega}}{2 t} \right) \sigma^{\dagger} R_{K_1} \left( \frac{- i\boldsymbol{\omega}}{2 t} \right)} \right]   + \frac{1}{2} \log \det \left[ \frac{1- r^2_2 R_{L_2} \left( \frac{- i\boldsymbol{\omega}}{2 t} \right)R_{K_2} \left( \frac{- i\boldsymbol{\omega}}{2 t} \right) }{1 - r^2_2 \sigma R_{L_2} \left( \frac{-i\boldsymbol{\Omega}}{2 t} \right) \sigma^{\dagger} R_{K_2} \left( \frac{- i\boldsymbol{\omega}}{2 t} \right)} \right] .
	\label{eq:SI:S2island}
\end{eqnarray}
Note that with trivial modifications, the last line is simply the entanglement between two 1+1D Majorana lattice with (possibly different) uniform nearest-neighbor hoppings.

\section{Derivation of transmission and reflection in the SYK defect CFT}
\label{append:SYK_transmission}

\subsection{Defect CFT}
Here we derive the transmission and reflection coefficients for the SYK$_2$ defect CFT. 
We first carry out coarse graining to obtain a continuum theory with the SYK defect, and then calculate the transmission and reflection coefficient from the two-point function of the stress tensor across the defect.
Consider the free Majorana chain,
\begin{eqnarray}
	H = i 2 t \sum_r \psi_r \psi_{r+1}, \quad \{\psi_r, \psi_r' \} = \delta_{ij}. 
\end{eqnarray}
After a Fourier transformation, it is easy to see that the Fermi momenta are located at $k_F = 0$ and $k_F = \frac{\pi}{a}$, where $a$ denotes the lattice constant.
Then, we take the following identification between the lattice fermion and the continuum fermion field  
\begin{equation}
	\begin{cases} 
		\psi_{2r} = \sqrt{\frac{a}{2}}[\psi_R(x_r) + \psi_L(x_r)], \\
		\psi_{2r+1} = \sqrt\frac{a}2[\psi_R(x_r) - \psi_L(x_r)],
	\end{cases} \quad r\in\mathbb{Z}, \quad x_r \equiv 2 r a.
\end{equation}
Here, $\psi_{R,L}(x)$ is the continuum fermion field operator with $\{\psi_L(x),\psi_L(x')\}=\{\psi_R(x),\psi_R(x')\}=\delta(x-x')$, and $\{\psi_L(x),\psi_R(x')\}=0$. Thus, the unit cell is effectively doubled with the two low-energy chiral modes folded to $k=0$. Note also that $\psi^{\dagger}_L(x)=\psi_L(x)$ and $\psi^{\dagger}_R(x)=\psi_R(x)$.

In terms of the continuum field, with $v_F \coloneq at$, the Hamiltonian becomes
\begin{eqnarray}
	H %
	&=& i \int \mathd x \, v_F (\psi_R \partial_x \psi_R - \psi_L \partial_x \psi_L).
\end{eqnarray}

For $N$ decoupled Majorana chains, we can duplicate the theory with an additional index $i=1,...,N$. 
Then we are ready to add the SYK interaction to the theory.
Without loss of generality, consider the SYK site to be the even site at $0$,
\begin{eqnarray}
	\sum_{jl} i J_{jl} \delta_{r,0} \psi^{\dagger}_{r,j} \psi_{r,l} =  \sum_{jl} i J_{jl} a \delta(x) [\psi_{j,R}(x) + \psi_{j,L}(x)][\psi_{l,R}(x) + \psi_{l,L}(x)] = \sum_{jl} i 2 a \delta(x) \tilde{J}_{jl} \Psi^{\dagger}_j(x) P \Psi_l(x),
\end{eqnarray}
where $\Psi=(\psi_L,\psi_R)^T$, 
$
P = \frac12 \begin{psmallmatrix}
	1 & 1 \\
	1 & 1
\end{psmallmatrix},
$
$\delta(x) = \delta(2ra) = \delta_{r,0}/2a$, and we have introduced the dimensionless coupling $\tilde{J}_{jl}=a J_{ij}$.
Recall that $J_{ij}$ is a Gaussian random variable with mean zero and variance given by~\eqref{seq:variance}.

Combined with the free part, it leads to the following continuum theory
\begin{eqnarray} \label{seq:lagrangian_SYK_defect}
	\mathcal L_E = \sum_j \Psi_j^\dag(x) (\partial_\tau -  i v_F \partial_x \sigma^z) \Psi_j(x) + \sum_{jl} i \delta(x) \tilde{J}_{jl} \Psi_j^{\dagger}(x) P \Psi_l(x).
\end{eqnarray}
Denoting $G \coloneq \frac1N \sum_i \langle \Psi_i \Psi_i^\dag \rangle $, the large-$N$ equation of motion reads
\begin{eqnarray}
	G(x,x'; \omega) &=& G_0(x,x'; \omega) + G_0(x,0; \omega) \Sigma(\omega) G(0,x'; \omega), \nonumber\\
	\Sigma(\omega) &=& 2\tilde J^2 P G(0,0;\omega) P.
\end{eqnarray}
where $\tilde J = J a$ with $J$ being the strength of the variance defined in~\eqref{seq:variance}.
In the following, we set $v_F = 1$ for simplicity; in other words, $a= 1/t$ and $\tilde J = J/t$. 
Notice that we have performed disorder average to derive the equation of motion.
The equation of motion is presented in mixed coordinates because the spatial translation symmetry is broken by the defect, whereas the temporal one is respected.
This large-$N$ equation of motion can be solved straightforwardly, leading to the solution:
\begin{eqnarray}
	G(x,y;\omega) &= &G_0(x,y; \omega) + \frac{i \sgn(\omega)(1+\tilde J^2 - \sqrt{1 + 2 \tilde J^2})}{4 \tilde J^2} \nonumber \\
	& &\times 
	\left( \begin{array}{cc}
		[\sgn(x) + \sgn(\omega)][\sgn(y) - \sgn(\omega)] &  -[\sgn(x) + \sgn(\omega)][\sgn(y) + \sgn(\omega)] \\
		-[\sgn(x) - \sgn(\omega)][\sgn(y) - \sgn(\omega)] & [\sgn(x) - \sgn(\omega)][\sgn(y) + \sgn(\omega)]
	\end{array} \right) e^{-(|x|+|y|) |\omega|} \\
	G_0(x,y;\omega) &=& i \left( \frac{\sgn(\omega)}2 + \frac{\sgn(x-y) \sigma^z}2 \right) e^{- |x-y| |\omega|}, \label{seq:G0}
\end{eqnarray}
which leads to the equal time correlation function 
\begin{eqnarray}
	G(x,y) \equiv G(x,y;\tau=0) = \frac{i\sigma^z}{x-y} + \frac{i(1+\tilde J^2 - \sqrt{1 + 2 \tilde J^2})}{2 \tilde J^2 (|x|+|y|)} \left( \begin{array}{cc}
		-\sgn(x) + \sgn(y) &  -\sgn(x) - \sgn(y)  \\
		\sgn(x) + \sgn(y) & \sgn(x) - \sgn(y)
	\end{array} \right).
\end{eqnarray}

The stress-energy tensor away from the defect is 
\begin{eqnarray}
	T(x) = 2 \psi^\dag(x) P_R \partial_x \psi(x), \quad \bar T = -2 \psi^\dag(x) P_L \partial_x \psi(x),
\end{eqnarray}
where $P_L = \frac12 (1 + \sigma^z)$, and $P_R = \frac12 (1 - \sigma^z)$ are the projection to right and left movers, respectively.
Using the full Green's function, the correlation function of stress-energy tensor is given by 
\begin{eqnarray}
	&& \langle T(x) T(-x) \rangle = -4 \Tr\left[ P_R  G^{(1,0)}(x,-x) P_R  G^{(1,0)}(x, -x)\right] = \frac{\left(1-\sqrt{1+2 \tilde J^2} \right)^2}{4\tilde J^4 (2x)^4}, \\
	&& \langle \bar T(x) \bar T(-x) \rangle = -4 \Tr\left[ P_L  G^{(1,0)}(x,-x) P_L  G^{(1,0)}(x, -x)\right] = \frac{\left(1-\sqrt{1+2 \tilde J^2} \right)^2}{4\tilde J^4(2x)^4}, \\
	&& \langle T(x) \bar T(x) \rangle = 4 \Tr\left[ P_R  G^{(1,0)}(x, x) P_L  G^{(1,0)}(x, x)\right] = \frac{\left(1 + \tilde J^2 - \sqrt{1+2 \tilde J^2} \right)^2}{4\tilde J^4(2x)^4}, \\
	&& \langle \bar T(-x) T(-x) \rangle = 4 \Tr\left[ P_L G^{(1,0)}(-x, -x) P_L  G^{(1,0)}(-x, -x)\right] = \frac{\left(1 + \tilde J^2 - \sqrt{1+2 \tilde J^2} \right)^2}{4\tilde J^4(2x)^4},
\end{eqnarray}
in which $G^{(1,0)}(x,y) \equiv \partial_x G(x,y)$ and $G^{(0,1)}(x,y) \equiv \partial_y G(x,y)$.
Hence, the transmission and reflection coefficients are \cite{quella2007reflection}
\begin{eqnarray}
	\mathcal T &=& \frac{\langle  T(x) T(-x) \rangle + \langle \bar T(x) \bar T(-x) \rangle }{ \langle  T(x) T(-x) \rangle + \langle \bar T(x) \bar T(-x) \rangle + \langle T(x) \bar T(x) \rangle + \langle \bar T(-x) T(-x) \rangle} = \frac2{3+\tilde J^2 - \sqrt{1+2 \tilde J^2}}, \label{eq:ss:T} \\
	\mathcal R &=& \frac{ \langle T(x) \bar T(x) \rangle + \langle \bar T(-x) T(-x) \rangle}{ \langle  T(x) T(-x) \rangle + \langle \bar T(x) \bar T(-x) \rangle + \langle T(x) \bar T(x) \rangle + \langle \bar T(-x) T(-x) \rangle} = 1-\frac2{3+\tilde J^2 - \sqrt{1+2 \tilde J^2}},
\end{eqnarray}
leading to the transmission coefficient presented in the main text. Since the conformal solution is continuously connected to that of a free Ising CFT, it also reveals that our defect formally corresponds to the Dirichlet boundary condition in the Ising CFT. The defect state here is $|  D,\phi \rrangle$ with $\phi = \sin^{-1}(\sqrt{\mathcal{T}})/2$~\cite{oshikawa1997boundary,brehm2015entanglement}.

\subsection{Lattice model}
\label{append:SYK-chain-g}

For completeness, we also derive the Green's function and transmission coefficient of the lattice model, which is infinite Majorana chains coupled at an SYK$_2$ site. The argument of the Green's function will follow the format $G(r,r',t;k,k',\omega)$ where the prime differentiates incoming and outgoing coordinates, and the semicolon separates spacetime and momentum-frequency coordinates, which are not present at the same time. The imaginary-time Dyson equation on the lattice reads
\begin{equation}
	G (r, r' ; i\omega_n) = G_0  (r, r' ; i\omega_n) + G_0  (r, 0 ; i\omega_n)
	\Sigma (\omega_n) G (0, r' ; i\omega_n),
\end{equation}
Set the lattice constant $a=1$. On a chain of length $L\to\infty$, we define the spatial Fourier transform
\begin{equation}
	G (k, k') \coloneq \frac{1}{L}\sum_{r,r'} G (r, r') e^{- i k r + ik' r'} .
\end{equation}
The free Green's function $G_0$ has the translational invariance $G_0 (r, r') = G_0 (r - r')$. Let
\begin{equation}
	G_0 (k) = \frac{1}{L}\sum_{\Delta r} G_0 {(\Delta r) }  e^{- ik \Delta r} = G_0 (0 ; k) = G_0  (0 ; k') .
\end{equation}
Then, for $m\in\mathbb{Z}$
\begin{equation}
	G_0 (k, k') = 2 \pi \delta (k - k' + 2m\pi) G_0 (k) .
\end{equation}
Recall that
\begin{eqnarray}
	G_0 (k, i\omega_n) & = & \frac{1}{-i \omega_n - 2 t \sin k}, \nonumber\\
	G_0 (0, 0 ; i\omega_n) & = & \frac{1}{\sqrt{-i \omega_n + 2 t}  \sqrt{-i
			\omega_n - 2 t}} = \frac{1}{\sqrt{- \omega^2_n - 4 t^2}} . 
\end{eqnarray}
In addition, we can obtain $\Sigma (\omega_n)$ using Eq.~\eqref{eq:inf} and \eqref{eq:Sigq},
\begin{equation}
	\Sigma (\omega_n) = \frac{i}{2}  \left[ \sqrt{- \omega_n^2 - 4 t^2} - \sqrt{- \omega_n^2 - 4 t^2 - 8 J^2} \right] .
\end{equation}
Solving $G (0 ; k', i\omega_n) = G_0  (0 ; k', i\omega_n) + G_0  (0, 0 ; i\omega_n) \Sigma (\omega_n) G (0 ; k', i\omega_n)$ for $G (0 ; k', i\omega_n)$ gives
\begin{eqnarray}
	G (0, k' ; i\omega_n) & = & \frac{1}{(-i \omega_n - 2 t \sin k')} \frac{2}{ 2 - i + i \sqrt{1 + 8 J^2 / (\omega^2_n + 4 t^2)} } . 
\end{eqnarray}
Solving $G (k, k', i\omega_n) = G_0  (k, k', i\omega_n) + G_0  (0; k, i\omega_n)
\Sigma (\omega_n) G (0 ; k', i\omega_n)$ for $G (k, k', i\omega_n)$ gives
\begin{eqnarray}
	G (k, k', i\omega_n) & = & 2 \pi \delta (k - k'+ 2m\pi) \frac{1}{-i \omega_n - 2 t
		\sin k} + \nonumber\\
	&  & \frac{1}{-i \omega_n - 2 t \sin k}  \sqrt{- \omega_n^2 - 4 t^2} 
	\frac{1-\sqrt{1 + 8 J^2 / (\omega^2_n + 4 t^2)}}{\sqrt{1 + 8 J^2 / (\omega^2_n + 4 t^2)} - 1 - 2 i } 
	\frac{1}{-i \omega_n - 2 t \sin k'} . 
\end{eqnarray}
Note that $G$ takes the standard form $G(z) = G_0(z) + G_0(z) \mathrm{T}(z) G_0(z)$ with the second term containing the T-matrix.%
For real frequencies, one substitute $\omega_n^2 \rightarrow - \omega^2$. The initial and final scattering states are on shell with respect to the free Hamiltonian, conforming to Fermi's golden rule. When on shell, $\omega^2 - 4 t^2 \rightarrow - 4 t^2 \cos^2 k$. Owing to energy conservation, at infinite time the surviving term is proportional to $\delta(2t \sin k - 2t \sin k')= [\delta(k - k'+ 2m\pi)+\delta(k + k' - \pi + 2m\pi)] / |2t \cos k|$. Thus, for an incoming state of momentum $k$, the outgoing momentum is $k$ and $\pi-k$.  The amplitude of the $\sqrt{\omega^2 - 4 t^2}$ factor also cancels the $\delta$-function measure $|2t \cos k|$ when on shell. Finally, the reflection coefficient is simply given by $\left|\frac{1-\sqrt{1 + 2 \sec^2(k) J^2 / t^2}}{\sqrt{1 + 2 \sec^2(k) J^2 / t^2} - 1 - 2 i}\right|^2$.

Let $\tilde{J}=J/t$. The transmission and reflection coefficients are
\begin{eqnarray}
	\mathcal{T}(k) &\coloneq& \left|\braket{k,t=\infty}{k,t=-\infty}\right|^2 =  \frac2{3+\tilde J^2 \sec^2 k - \sqrt{1+2 \tilde J^2 \sec^2 k}}, \nonumber\\
	\mathcal{R}(k) &\coloneq& \left|\braket{\pi-k,t=\infty}{k,t=-\infty}\right|^2 = 1-\frac2{3+\tilde J^2 \sec^2 k - \sqrt{1+2 \tilde J^2 \sec^2 k}}.
\end{eqnarray}
In the low-energy limit, $k \sim 0$. Then we recover the conformal result in Eq.~\eqref{eq:T} or \eqref{eq:ss:T}. Hence, there is a simple relation between the defect strengths in the continuum and on a lattice. These results may also be obtained by the LSZ formalism in the continuum limit~\cite{delfino1994scattering}.

\section{Discussions on R\'enyi entropy and transmission coefficient}
\label{append:discussion_renyi}

\subsection{Derivation of the transmission coefficient for energy defect in the Ising CFT}

In this section, we briefly review the calculation of the transmission coefficient for an energy defect in the Ising CFT, i.e., the free Majorana model.
We used a similar model as in the previous section to evaluate the transmission and reflection coefficient.
In the presence of an energy defect, the continuum field theory reads
\begin{eqnarray}
	\mathcal L_E= \Psi^\dag(x) (\partial_\tau - i \sigma^z \partial_x) \Psi(x) + g \delta(x) \Psi^\dag(x) \sigma^y \Psi(x),
\end{eqnarray}
where $g$ denotes the defect strength, and we set the Fermi velocity to be one for simplicity. 
Comparing this Lagrangian with the SYK defect~\eqref{seq:lagrangian_SYK_defect}, the defect has a different form, and because the energy defect does not couple different flavors, we consider $N=1$.

The scattering of the energy defect leads to the Schwinger-Dyson equation,
\begin{eqnarray}
	G(x_1, x_2; \omega) &=& G_0(x_1 - x_2; \omega) + G_0(x_1 - x_2; \omega) \Sigma(\omega) G_0(x_2; \omega) \\
	\Sigma(\omega) &=& g \frac{\sigma^y}{1-g G_0(0;\omega) \sigma^y}.
\end{eqnarray}
where $G = \langle \Psi \Psi^\dag \rangle$ is the full propagator, and $G_0(x-y;\omega) = G_0(x,y;\omega)$ in~\eqref{seq:G0}. 
Note that this Schwinger-Dyson equation is exact. 
The solution is
\begin{eqnarray}
	G(x_1, x_2, \tau_1, \tau_2) =  
	\left( \begin{array}{cc}
		\frac{i[4+g^2 - 2g^2(\Theta(x_1) \Theta(-x_2) + \Theta(-x_1) \Theta(x_2))]}{2\pi(4+g^2)(x_{12} + i \tau_{12}))} 
		&  \frac{i 2g (\Theta(x_1) \Theta(x_2) - \Theta(-x_1) \Theta(-x_2))}{\pi(4+g^2)(x_1 + x_2 + i \tau_{12}))} \\
		-\frac{i 2g (\Theta(x_1) \Theta(x_2) - \Theta(-x_1) \Theta(-x_2))}{\pi(4+g^2)(x_1 + x_2 + i \tau_{12}))} 
		& -\frac{i[4+g^2 - 2g^2(\Theta(x_1) \Theta(-x_2) + \Theta(-x_1) \Theta(x_2))]}{2\pi(4+g^2)(x_{12} + i \tau_{12}))}
	\end{array} \right) ,
\end{eqnarray}
where $x_{12} \coloneq x_1 - x_2$ and $\tau_{12} \coloneq \tau_1 - \tau_2$

With the full propagator, we can evaluate the transmission and reflection coefficients via stress-energy tensors similar to the preceding section.
The results are
\begin{eqnarray} \label{seq:transmission}
	\mathcal{T} = 1 - \frac{16g^2}{(4+g^2)^2}, \quad \mathcal{R} = \frac{16g^2}{(4+g^2)^2}.
\end{eqnarray}
This is consistent with Ref.~\onlinecite{delfino1994scattering}.

In comparison, we note that the transmission and reflection coefficients in the bosonic counterpart is $\sqrt{\mathcal{T}}=1-\frac{ig}{4\omega+ig}$ and $\sqrt{\mathcal{R}}=\frac{-ig}{4\omega+ig}$~\cite{delfino1994scattering}. At low frequency, the defect is relevant, and accordingly $\mathcal{T}=0$. This is discussed in detail later in Sec.~\ref{sec:bsyk}.

\subsection{Relation between the effective central charge and the transmission coefficient}

We discuss the relation between the transmission coefficient and the entanglement entropy. 
The entanglement entropy across the defect is given by an effective central charge $\tilde c_n$ \cite{calabrese2009entanglement}
\begin{equation}
	S_n = \frac{\tilde c_n (\mathcal T)}{12}\left( 1+\frac{1}{n} \right) \log \left( \frac{L}{a} \right),
\end{equation}
where $S_n$ denotes R\'enyi-$n$ entropy and $a$ is the lattice constant.
For von Neumann entropy, the effective central charge is~\cite{eisler2010solution}
\begin{eqnarray}
	\tilde{c}_1 = - \frac{6}{\pi^2} \left\{ \left[ (1+s) \log (1+s) + (1-s) \log(1-s) \right] \log s + (1+s) \mathrm{Li}_2(-s) + (1-s) \mathrm{Li}_s(s) \right\}, \quad s = \sqrt{\mathcal{T}}.
\end{eqnarray}
For the R\'enyi entropy, the results can be found in~\cite{peschel2012exact}. 
In particular, the R\'enyi-2 entropy calculated in our paper is 
\begin{eqnarray}
	\tilde c_2(\mathcal T) = \frac{8}{\pi^2} \arcsin^2\left(\sqrt{\frac{\mathcal T}{ 2}}\right).
\end{eqnarray}

For the energy defect in the Majorana chain model, the transmission coefficient reads~\cite{eisler2010solution} $ \sqrt{ \mathcal{T} } = \frac2{1+ g +1/(1+g)} $, where $g$ is a parameter that captures the strength of the defect.
In particular, for the Majorana chain model with a modified bond hopping studied in the main text, $g = (t'-t)/t$, with $t$ ($t'$) denoting the normal (defect) bond hopping strength.
Note that the relation between the transmission coefficient derived from the continuum field theory~\eqref{seq:transmission}, and from the transfer matrix method is not straightforward. 
Nevertheless, they agree at the leading order as expected, since the continuum theory neglects higher-order contribution:
\begin{eqnarray}
	\mathcal T \approx 1 - g^2,
\end{eqnarray}
Therefore, the field theory calculation correctly predicts the leading behavior of the effective central charge 
\begin{eqnarray}
	\tilde c_2 \approx \frac1{4\pi^2} - \frac{g^2}{2\pi^2}. 
\end{eqnarray}
However, such a minimal consistency is violated for the SYK defect as discussed in the main text, as will be shown next.

\subsection{Perturbative expansions for the SYK defect CFT}

In the interface geometry where $L_1=K_2=0$, the R\'enyi-2 entropy in \eqref{eq:SI:S2island} or \eqref{eq:S2island} in the main text simplifies to
\begin{equation}
	S_2/N = \frac{1}{2} \Tr \log [G_{(2)} \tilde{G}^{-1}] + J^2  \left( \frac{1}{4 q} - \frac{1}{4} \right) 
	\int \mathd \tau_1 \mathd \tau_2  \{ [2 \tilde{G} (\tau_1, \tau_2)]^q - [2 G_{(2)} (\tau_1, \tau_2)]^q \}.
\end{equation}
To leading order in $J$ at $q=2$,
\begin{eqnarray}
	\frac{1}{2} \Tr \log [G_{(2)} \tilde{G}^{-1}] &=& - \frac{1}{2} \Tr \log [G_{(2)}^{- 1} \tilde{G}] + J^2  \left( \frac{1}{4 q} - \frac{1}{4} \right) 
	\int \mathd \tau_1 \mathd \tau_2  \{ [2 \tilde{G} (\tau_1, \tau_2)]^q - [2 G_{(2)} (\tau_1, \tau_2)]^q \}.
\end{eqnarray}
Let $\Gamma_{(2)} \equiv \left. G_{(2)} \right|_{J = 0} = \left(- i\tmmathbf{\omega}-
t\mathbf{D}_{K_1, L_1} - t\mathbf{D}_{K_2, L_2}\right)^{- 1}$ and $\tilde{\Gamma}
\equiv \left. \tilde{G} \right|_{J = 0} = \left(- i\tmmathbf{\omega}-
t\mathbf{D}'_{K_1, L_1} - t\mathbf{D}'_{K_2, L_2}\right)^{- 1}$. Then
\begin{eqnarray}
	\frac{1}{2} \tmop{Tr} \log \left[ G_{(2)} \tilde{G}^{-1}  \right] & = &
	\frac{1}{2} \tmop{Tr} \log \left[ \frac{- i\tmmathbf{\omega}- 2 J^2 
		\tilde{\Gamma} - t\mathbf{D}'_{K_1, L_1} - t\mathbf{D}'_{K_2, L_2}}{-
		i\tmmathbf{\omega}- 2 J^2 \Gamma_{(2)} - t\mathbf{D}_{K_1, L_1} -
		t\mathbf{D}_{K_2, L_2}}  \right] \nonumber\\
	& = & \frac{1}{2} \tmop{Tr} \log \left[ \frac{- i\tmmathbf{\omega}- 2 J^2 
		\tilde{\Gamma} - tR_{K_1} \left( \tfrac{- i\tmmathbf{\omega}}{2 t} \right) -
		tR_{L_2} \left( \tfrac{- i\tmmathbf{\omega}}{2 t} \right)}{-
		i\tmmathbf{\omega}- 2 J^2 \Gamma_{(2)} - tR_{K_1} \left( \tfrac{-
			i\tmmathbf{\omega}}{2 t} \right) - t \sigma R_{L_2} \left( \tfrac{-
			i\tmmathbf{\omega}}{2 t} \right) \sigma^{\dag}}  \right] \label{eq:ss:icft-sRs}\\
	& = & \frac{1}{2} \tmop{Tr} \log \left[ G_{(2)} \tilde{G}^{-1}  (1 + 2
	J^2 \Gamma_{(2)} G_{(2)} - 2 J^2  \tilde{\Gamma}  \tilde{G}) \right] + O
	(J^4) \nonumber\\
	& = & \frac{1}{2} \tmop{Tr} \log [\tilde{\Gamma}^{- 1} \Gamma_{(2)} ] + J^2
	\tmop{Tr} [\Gamma^2_{(2)} - \tilde{\Gamma}^2 ] + O (J^4) . 
\end{eqnarray}
For the integral,
\begin{equation}
	- \frac{3 J^2}{4} \int \mathd \tau_1 \mathd \tau_2  \{ [\tilde{G} (\tau_1,
	\tau_2)]^2 - [G_{(2)} (\tau_1, \tau_2)]^2 \} = - \frac{3 J^2}{4} \int \mathd
	\tau_1 \mathd \tau_2  \{[\tilde{\Gamma} (\tau_1, \tau_2)]^2 -
	[\Gamma_{(2)} (\tau_1, \tau_2)]^2\} .
\end{equation}
Due to fermionicity, $\Gamma (\tau_1, \tau_2) = - \Gamma (\tau_2, \tau_1)$. Then $\tmop{Tr}
(\Gamma^2) \equiv \int \mathd \tau_1 \mathd \tau_2 \Gamma (\tau_1, \tau_2)
\Gamma (\tau_2, \tau_1) = - \int \mathd \tau_1 \mathd \tau_2 \Gamma (\tau_1,
\tau_2)^2$. Thus,
\begin{equation}
	\frac{S_2}{N} = \left. \frac{S_2}{N} \right|_{J = 0} + \frac{J^2}{4} \int
	\mathd \tau_1 \mathd \tau_2  \{[\tilde{\Gamma} (\tau_1, \tau_2)]^2 -
	[\Gamma_{(2)} (\tau_1, \tau_2)]^2\} + O (J^4) .
\end{equation}
Therefore, the leading order change in $S_2$ and hence $\tilde{c}_2$ is $J^2$.
However, from Eq.~\eqref{eq:ss:T} or \eqref{eq:T} in the main text, $\mathcal{T}=2/\left(3+\tilde{J}^2-\sqrt{1+2\tilde{J}^2}\right)\approx1-J^4/2$. Consequently, the leading-order change in $\tilde{c}_2(\mathcal{T})$ is $J^4$, in disagreement with that extracted from the scaling of $S_2$, as plotted in Fig.~\ref{fig:c2}(d) in the main text.

Finally, we address the independence of the $g$-function on $J$ in the thermodynamic limit. In the setup to compute the $g$-function, $K_1=K_2=K$ and $L_1=L_2=L$. When $K$ is large, $\lim_{K\to\infty}\mathbf{D}_{K,L}=\lim_{K\to\infty}D'_{K,L}(\tfrac{-i\boldsymbol{\omega}}{2t})=\lim_{K\to\infty}R_{K+L}(\tfrac{-i\boldsymbol{\omega}}{2t})$ for low-frequency components. In this case, the defect is too far from the interface to affect each other. Thus, $G_{(2)}\approx \tilde{G}$. As a result, $S_2$ comes only from noninteracting chains. In contrast, when the defect is next to the interface, e.g., $K_2=0$ as in \eqref{eq:ss:icft-sRs}, the twist operator is immediately present in $G_{(2)}$, so its effect survives in the thermodynamic limit.

\section{Further applications}

In this section, we discuss further application of our saddle-point numerical functional determinant method.

\subsection{Bosonic SYK defect}\label{sec:bsyk}

\begin{figure}
	\includegraphics[width=0.55\linewidth]{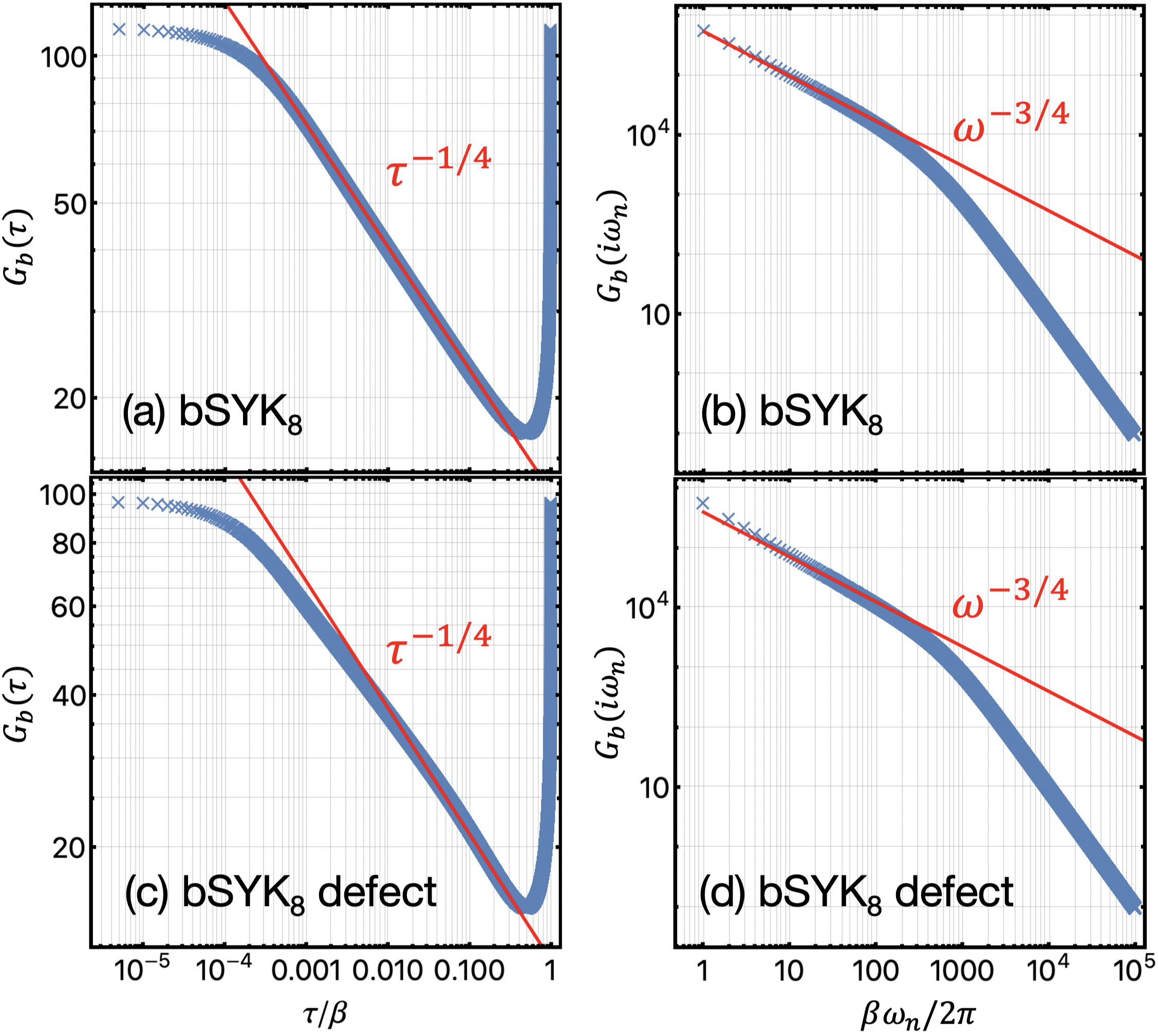}
	\caption{\label{fig:ss:bsyk}Green's function in imaginary time and Matsubara frequency of (a)--(b) the
		bosonic SYK$_8$ model, and (c)--(d) the bosonic SYK$_8$ defect coupled to a
		large $N$ number of chains \eqref{seq:dyson-bsyk}. Red lines show the corresponding power laws of
		the conformal solution \eqref{seq:bSYKconformal}. The parameters are $J / t = 1$, $\beta J = 2000$.}
\end{figure}

Our method can be easily extended to other defect theories. An example is the Yukawa-SYK defect %
coupling $N$ free majorana channels and $M$ free boson channels at the defect site~\cite{wang2020ysyk}.
Another example is $N$ bosonic channels coupled at a bosonic SYK (bSYK) defect. 
We showcase the bSYK defect model below, and leave the study of the Yukawa-SYK defect for future work~\cite{ge2024ysyk}. 

Consider the bosonic counterpart to our model. The action reads
\begin{equation}
	- I = - \frac{1}{2}  \sum_{j, r} \int d \tau [(\partial_{\tau} \varphi_{j,
		r})^2 - 2 t \varphi_{j, r} \varphi_{j, r + 1} + 2 t \varphi_{j, r}^2] + \int
	d \tau \sum_{j_1, ..., j_q} J_{j_1, ..., j_q} \varphi_{j_1, 0} ...
	\varphi_{j_q, 0} .
\end{equation}
The chain index is $j = 1, \ldots, N$. To the left of the defect are $L$
sites, and to the right $K$ sites. In the large $N$ limit, assuming time translational invariance, the
Schwinger-Dyson equations in imaginary time $\tau$ and Matsubara frequencies
$\omega_n = 2 \pi n / \beta$ reads
\begin{eqnarray}
	\label{seq:dyson-bsyk}
	G_b (i \omega_n) & = & \left[ \omega^2_n + 2 t - tR_L \left(
	\frac{\omega_n^2}{2 t} + 1 \right) - tR_K \left( \frac{\omega_n^2}{2 t} + 1
	\right) - \Sigma_b (i \omega_n) \right]^{- 1}, \nonumber\\
	\Sigma_b (\tau_1, \tau_2) & = & J^2 G_b (\tau_1, \tau_2)^{q - 1} .
\end{eqnarray}
For $t = 0$ we recover the self-consistence equations for the bSYK.
Albeit its unstable nature, one can solve the IR equations formally,
\begin{equation}
	G^c_b(i\omega) = -1/\Sigma^c_b(i\omega), \qquad 
	\Sigma^c_b(\tau) = J^2 G^c_b(\tau)^{q-1},
\end{equation}
for a conformal solution $G^c_b$ when $q\ne2$~\cite{murugan2017more},
\begin{eqnarray}
	G^c_b(\tau) &=& \left[ \left(\frac{1}{2}-\frac{1}{q}\right) \frac{\cot(\pi/q)}{\pi J^2}\right]^{1/q} \left| \tau \right|^{-2/q}, \nonumber \\
	\Sigma^c_b(i\omega) &=& \left[ \left(\frac{1}{2}-\frac{1}{q}\right) \frac{\cot(\pi/q)}{\pi J^2}\right]^{-1/q} \frac{\left|\omega\right|^{1-2/q}}{2 \sin(\pi/q) \Gamma(1-2/q)}.
\end{eqnarray}
At finite temperatures,
\begin{eqnarray}
	G^c_b (\tau) & = & \left[ \left( \frac{1}{2} - \frac{1}{q} \right)
	\frac{\pi \cot (\pi / q)}{\beta^2 J^2} \right]^{1 / q} \left| \sin \left(
	\frac{\pi \tau}{\beta} \right) \right|^{- 2 / q}, \nonumber\\
	\Sigma^c_b (i \omega_n) & = & \frac{\beta J^{2 / q}}{\pi} \left[
	\frac{\beta^2}{2 \pi} \frac{\tan (\pi / q)}{1 - 2 / q} \right]^{- 1 + 1 / q}
	\sin \left( \frac{\pi}{q} \right) \mathrm{B} \left( \frac{2}{q} - 1, n+1 - \frac{1}{q}\right) , \label{seq:bSYKconformal}
\end{eqnarray}
where $\mathrm{B}(x,y)\coloneq \Gamma(x)\Gamma(y)/\Gamma(x+y)$ is the beta function. Since the engineering dimension $[\varphi (x, \tau)] = 0$, the SYK interaction is relevant for all $q$'s. %
Consequently, the conformal solutions above hold at IR for the defect as well. However, since the solution is unstable, some pinning is required to converge to this solution. 
Below, we outline the intricacies in the numerics.

We discretize the imaginary time into $\delta \tau$, and truncate the Matsubara frequency, i.e., $- \beta / 2 \delta\tau < n \leqslant \beta / 2 \delta \tau$, so that the time and frequency grids are the same size. 
Since bSYK itself is unstable \cite{murugan2017more,shen2023long}, we enforce both time translational invariance and particle-hole symmetry, i.e., $G_b (\tau_1, \tau_2) = G_b (\tau_1 - \tau_2)$, and $G_b (\tau_1 - \tau_2) = G_b (\tau_2 - \tau_1)$.
Furthermore, we pin the zero frequency term of $G_b^{-1}$ to the conformal solution \cite{tulipman2020strongly}, and instead iterate with
${G'_b}^{- 1} (i \omega_n) \coloneq G_b^{- 1} (i \omega_n) + \Sigma_b (0) - \Sigma^c_b (0) - \Sigma_b (i \omega_n)$.
The result is shown in Fig.~\ref{fig:ss:bsyk}. 
It is clear that the SYK term dominates IR physics. 
Note that although the power law agrees with the conformal solutions, the exact values are not identical. 
In particular, for the converged solution $\Sigma_b (0) \ne \Sigma^c_b (0)$.

In the bosonic case, the transmission is completely cut off for the SYK defect, as it is for a mass defect. Both can be solved in the LSZ formalism. 
Recall that the impurity $T$-matrix satisfy a Dyson series made up of the bare local Green's function $g_0(\omega+i\eta) \coloneq G_0(x=y=0,\omega+i\eta)$, and the bare self-energy at the impurity site $\Sigma_0$~\cite{delfino1994scattering},
\begin{equation}
	T(\omega+i\eta) = \frac{1}{\Sigma_0^{-1}(\omega+i\eta)-g_0(\omega+i\eta)}.
\end{equation} 
In the case of a mass defect of the form $\int d\tau dx J^2\phi^2(x) \delta(x)$, the bare self-energy $\Sigma_0(\omega+i\eta)=i J^2$. The reflection coefficient can be obtained from the LSZ approach following Ref.~\onlinecite{delfino1994scattering}, using $\mathcal{R} = |T/2p|^2$ where $p$ is the momentum of the on-shell boson. It further simplifies to $\mathcal{R} = |T/2\omega|^2$ in our massless case. For the 1+1D real boson, $g_0(\omega+i\eta)=1/\sqrt{(\omega+i\eta)^2}$. The result for the mass defect is
\begin{equation}
	\label{seq:Rboson}
	\mathcal{R} = \left| \frac{iJ^2}{2\omega-iJ^2} \right|^2 \xrightarrow{\omega\to0} 1.
\end{equation}
This also follows from the classic solution to a delta potential scattering in 1D~\cite{lipkin1973quantum}.
As for the bSYK defect, let $\sigma(\tau)=g_0(\tau)^{q-1}$, then $\Sigma_0(\omega+i\eta)=iJ^2 \sigma(\omega+i\eta)$. Thus, one can substitute $J^2\to J^2 \sigma(\omega+i\eta)$ in \eqref{seq:Rboson} above. Since $[\phi(x,\tau)]=0$,  $[\sigma(\tau)]=0$, and $[\sigma(\omega+i\eta)]=-1$. Therefore, in the limit $\omega \to 0$, one can see that $\mathcal{R}= J^4 / [J^2+O(\omega^2)]^2 \to 1$, without explicit calculations.

\subsection{Interacting bulk CFT}

Here we discuss how to extend our saddle-point method in the presence of
onsite self-energy to accommodate interacting CFTs in the bulk. A concrete
example is the SYK lattice in 1+1D at large $N$, with a defect site indexed 0.
Once again we introduce the conjugate-pair fields $(G_r, \Sigma_r)$, but now
on each site $r$. Thus, $G_j$ becomes the local Green's function, i.e., $G_r
(\tau_1, \tau_2) \sim \psi_r (\tau_1) \psi_r (\tau_2) / N$, and $\Sigma_r$
becomes the self-energy on site $r$. The part of the action containing the
original field $\psi$ assumes the form
\begin{equation}
	-I[\psi] %
	= - \frac{1}{2}  \sum_{j, r, r'} \int d \tau_1 d \tau_2
	\, \psi_{j, r} (\tau_1)  [(\partial_{\tau_1} \delta_{r, r'} - i t h_{r, r'}) \delta (\tau_1 - \tau_2) - \Sigma_r (\tau_1, \tau_2)\delta_{r,r'}] \psi_{j, r'} (\tau_2),
\end{equation}
where $h$ is the hopping matrix. Denote $A=(\partial_{\tau_1} \delta_{r,
	r'} - i t h_{r, r'}) \delta (\tau_1 - \tau_2) - \Sigma_r (\tau_1, \tau_2)\delta_{r,
	r'}$.
This leads to as many saddle-point equations as the length of the lattice.
Denote the minor of $A$ with respect to entries $i, j$ by $[A]_{i j}$, the
saddle-point equations are
\begin{eqnarray}
	G_r  (i \omega_n) & = & \det [A]_{r r} (i \omega_n) / \det A (i \omega_n),
	\nonumber\\
	\Sigma_r & = & \Sigma_r [G_r] . 
\end{eqnarray}
The self-energy $\Sigma_r$ is a function of $G_r$, same as before. Therefore,
naively one needs to solve $2L$ coupled self-consistency equations, which limits the system size accessible.

A reduction in the finite-size effect can be achieved if the left and right ends of the system are coupled to chains of infinite length, acting as baths. They are assumed to have a translationally invariant self-energy. Then, the function determinant including both the defect region ($d$) and the baths to the left ($L$) and right ($R$) is given by
\begin{eqnarray}
	\frac12 \log \left( \det A_{L} \det A_{d} \det A_{R} \right), 
\end{eqnarray}
in which
\begin{eqnarray} \label{eq:ALR}
	&& A_{L/R} = - i \omega - \Sigma_{L/R} - \frac{t^2}{-i \omega - A_{L/R}}, \\
	&& A_d = A - \frac{t^2}{-i \omega - A_{L}} - \frac{t^2}{-i \omega - A_{R}}.
\end{eqnarray}
Here, the chains are taken to be infinite as follows.
A general iteration equation at the $n$-th left (or right) most site is 
\begin{eqnarray}
	A_n = - i\omega - \Sigma_n - \frac{t^2}{-i\omega - A_{n-1}}, \quad A_1 = -i\omega - \Sigma_1. 
\end{eqnarray}
Assuming a translationally invariant $\Sigma_n$ and taking $n \rightarrow \infty$ gives the above equation for $A_{L/R}$. 
With~\eqref{eq:ALR}, 
\begin{eqnarray}
	A_{L/R} = - i\omega - \frac12 \left( \Sigma_{L/R} + \sgn(\omega) \sqrt{4t^2 + \Sigma^2_{L/R}} \right)
\end{eqnarray}
The Dyson equation in the frequency space is
\begin{eqnarray}
	&& G_{L/R} =  - \left[ A_{L/R}^{-1}  + A_d^{-1} \left(\frac{t}{-i \omega - A_{L/R}} \right)^2 \right] \frac{\partial A_{L/R}}{\partial \Sigma_{L/R}}, \quad G_d = A_d^{-1}, \\
	&& \frac{\partial A_{L/R}}{\partial \Sigma_{L/R}} = - \frac12 \left( 1+ \frac{\sgn(\omega)\Sigma_{L/R}}{\sqrt{4t^2 + \Sigma_{L/R}^2} }\right).
\end{eqnarray}
The self-energies are again a function of the respective local Green's functions.

\subsection{Periodic boundary condition}

We provide a simple example to illustrate how our method can be extended to the periodic boundary condition. Let the system be divided into $A$, $B$, and $C$ blocks, the sizes of which are $K \times K$, $L \times L$, and $M \times M$, respectively. The defect is located in $B$. Define a shorthand for the special matrices $(e_{mn})_{ij}\coloneq\delta_{i,m}\delta_{n,j}$. On a periodic boundary condition with short-range hoppings $t$, the functional determinant takes the form
\begingroup
\newcommand{\td}{\ensuremath{t^\dagger}}
\begin{eqnarray}
	&&\det  \left(\begin{array}{ccccc}
		\multicolumn{2}{c|}{\multirow{2}{*}{\ensuremath{A}}}  &  &  & \ensuremath{t}\\ 
		&  \multicolumn{1}{c|}{} & \ensuremath{t^\dagger} &  & \\ \cline{1-3}
		& \multicolumn{1}{c|}{t} &  \ensuremath{B} & \multicolumn{1}{|c}{\ensuremath{t^\dagger}} & \\ \cline{3-5}
		&  & \ensuremath{t} & \multicolumn{2}{|c}{\multirow{2}{*}{\ensuremath{C}}} \\
		\ensuremath{t^\dagger} &  &  & \multicolumn{1}{|c}{} & 
	\end{array}\right) 
	= 
	\det A \det \left(\begin{array}{cccc}
		\multicolumn{2}{c}{\multirow{2}{*}{\ensuremath{B - e_{11} t A^{- 1}_{KK} t^\dagger}}} & \multicolumn{1}{|c}{} & \multicolumn{1}{r}{\ensuremath{t A^{- 1}_{K 1}t}} \\ 
		& &  \multicolumn{1}{|c}{\ensuremath{t^\dagger}}  \\ \cline{1-4}
		& \multicolumn{1}{r}{\ensuremath{t}} & \multicolumn{2}{|c}{\multirow{2}{*}{\ensuremath{C -  e_{MM} \td A_{11}^{- 1}t}}}  \\
		\ensuremath{t^\dagger A^{- 1}_{1 K}t^\dagger} &  \multicolumn{1}{r}{} &  \multicolumn{1}{|c}{} 
	\end{array}\right)
	\nonumber \\
	& = & \det A \det (C -  e_{MM} \td A_{11}^{- 1}t) \times  \det \Big[ B - e_{11} \td A^{- 1}_{KK} t - e_{KK} t A_{K1}^{-1} t (C - e_{MM} \td A_{11}^{- 1} t)^{- 1}_{11}  \td A_{1K}^{-1} \td   \nonumber\\
	& &  \  -e_{1L} t A^{- 1}_{K 1} t (C - e_{MM} \td A_{11}^{- 1} t)^{- 1}_{K 1} t  -e_{L1} \td (C -  e_{MM} \td A_{11}^{- 1} t)^{-1}_{1 K} \td A^{- 1}_{1 K} \td
	- e_{LL}\td (C -  e_{MM} \td A_{11}^{- 1} t)^{- 1}_{MM} t \Big]. \qquad
\end{eqnarray}
\endgroup
The last determinant can be easily computed if $B$ is a defect or a small region. In general, one will encounter the determinant of matrices of the following kind,
\begin{equation}
	A_1(h_1;\omega_1;\omega_2\dots\omega_n-1;\omega_n) \coloneq \begin{pmatrix}
		\omega_1 & t^{\dag} &  &  & h\\
		t & \omega_2 & \ddots &  & \\
		& \ddots & \ddots & \ddots & \\
		&  & \ddots & \ddots & t^{\dag}\\
		h^{\dag} &  &  & t & \omega_n
	\end{pmatrix}.
\end{equation}
The determinant of $A_1$ can be computed recursively:
\begin{eqnarray}
	\det A_1 (h_1 ; \omega_1 ; \omega_2 \ldots \omega_{n - 1} ; \omega_n) & = &
	\omega_1 \det A_2 (t \omega_1^{- 1} h_1 ; \omega_2 - t \omega_1^{- 1}
	t^{\dag} ; \omega_3 \ldots \omega_{n - 1} ; \omega_n - h_1^{\dag}
	\omega_1^{- 1} h_1), \nonumber\\
	\det A_{m - 1} {(h ; \nu ; \omega_m}  \ldots \omega_{n - 1} ; \lambda) & = &
	\nu \det A_m (t \nu^{- 1} h ; \omega_m - t \nu^{- 1} t^{\dag} ; \omega_{m
		+ 1} \ldots \omega_{n - 1} ; \lambda - h^{\dag} \nu^{- 1} h) . 
\end{eqnarray}
The process continues until $A_{n-2}$, when the matrix dimension is reduced to 3. This has the same order of complexity as Eq.~\eqref{eq:ss:Dgeneral}.

\end{document}